\documentclass[12pt]{article}
\usepackage{amsmath,amssymb}
\usepackage{bm}
\usepackage{mathrsfs}
\usepackage{fourier}
\usepackage{multirow}
\allowdisplaybreaks[4]
\newcommand{{\Slashp}}{p\!\!\!\!\!\big/}
\newcommand{{\Slashq}}{q\!\!\!\!\!\big/}
\newcommand{{\Slashk}}{k\!\!\!\!\!\big/}
\newcommand{\LBox}{\mbox{\large$\Box$}}
\usepackage[usenames]{color}

\begin{document}

\title{Fermionic scalar field}

\author{
Yoshiharu \textsc{Kawamura}\footnote{E-mail: haru@azusa.shinshu-u.ac.jp}\\
{\it Department of Physics, Shinshu University, }\\
{\it Matsumoto 390-8621, Japan}\\
}

\date{
June 24, 2014}

\maketitle
\begin{abstract}
We reexamine the connection between spin and statistics 
through the quantization of a complex scalar field,
using the formulation with the property
that the hermitian conjugate of canonical momentum for a variable is
just the canonical momentum for the hermitian conjugate of the variable.
Starting from an ordinary Lagrangian density
and imposing the anti-commutation relations on the field, 
we find that the difficulty stems from
not the ill-definiteness (or unboundedness) of the energy 
and the breakdown of the causality
but the appearance of states with negative norms.
It is overcome by introducing an ordinary scalar field
to form a doublet of fermionic symmetries, 
although the system becomes empty leaving the vacuum state alone.
These features also hold for the system with a spinor field
imposing the commutation relations on.
\end{abstract}

\newpage

\abovedisplayskip=1.0em
\belowdisplayskip=1.0em
\abovedisplayshortskip=0.5em
\belowdisplayshortskip=0.5em
 
\parskip=0.5em
\tableofcontents
\parskip=0.25em

\maketitle

\section{Introduction}
\label{Introduction}

{\it Observed particles of integer spin obey Bose-Einstein statistics
and are quantized using the commutation relations,
and those of half-odd-integer spin obey Fermi-Dirac statistics
and are quantized using the anti-commutation relations.}
These properties are explained under conditions 
such as the positivity of energy, the microscopic causality 
and the positive-definiteness of norm
in the framework of relativistic quantum field theory,
and is known as 
the spin-statistics theorem~\cite{P&B,deWet,P1,Feyn1,Feyn2,P2,Sch,L&Z,Bur,S&W,D&S}.

Although it is taken for granted that such a connection between spin and statistics 
exists in nature, it is not so clear what happens when particles are quantized by imposing
abnormal relations on.
There seems to be a confusion 
or the absence of common understanding
on which condition is essential to the theorem 
or incompatible with abnormal relations.
It might cause mainly from the difference of setup or preconditions,
and hence model-dependent analyses would be useful to avoid ambiguities
and to help a deeper understanding of spin and statistics.
From this viewpoint, we deal with explicit models at the expense of a generality,
taking a clue from the features of Faddeev-Popov ghost fields.

Faddeev-Popov ghost fields are hermitian scalar fields 
appearing on the quantization of gauge theories,
and follow the anti-commutation relations~\cite{F&P}.
They have positive energies and respect the causality 
but generate states with negative norms.
It is natural to expect that these features are shared in a wide class of models.
Actually, Fujikawa has studied the spin-statistics theorem in the path integral formalism,
with expressions of the operator formalism,
and pointed out that the causality is ensured regardless of statistics
and the positive norm condition is crucial to the theorem~\cite{F}.\footnote{
Pauli reconsidered the spin-statistics theorem using the formulation based on
the Fock space with indefinite metrics, and found that
the positive norm condition plays a vital role in the theorem~\cite{P2}.
}
Our study is regarded as a manifestation of the statement
based on explicit models in the operator formalism.

In this paper, we reexamine the connection between spin and statistics through
the quantization of a complex scalar field.
Starting from an ordinary Lagrangian density
and imposing the anti-commutation relations on the scalar field,
we find that the difficulty stems from
not the ill-definiteness (or unboundedness) of the energy 
and the breakdown of the causality
but the appearance of states with negative norms.
We refer to such an abnormal scalar field as a $\lq$fermionic scalar field'.
The difficulty is overcome by introducing an ordinary scalar field
to form a doublet of fermionic symmetries,
although the system becomes empty leaving the vacuum state alone.
These features also hold for the system with a spinor field
imposing the commutation relations on.
We refer to such an abnormal spinor field as a $\lq$bosonic spinor field'.
As a by-product, we construct analytical mechanics
in the form with the manifestly hermitian property
such that {\it the hermitian conjugate of canonical momentum for a variable is
just the canonical momentum for the hermitian conjugate of the variable.}

The contents of this paper are as follows.
In Sec. II, we study the system with harmonic oscillators
imposing the anti-commutation relations on variables, as a warm-up.
In Sec. III,  we examine the system of a fermionic scalar field 
and clarify the difficulty on quantization.
We investigate the system containing both an ordinary complex scalar field
and a fermionic one.
Section IV is devoted to conclusions and discussions.
Appendices also contains new ingredients.
In Appendix A, we present useful formulas of differentiation for variables and
a new definition of the canonical momenta, the Hamiltonian and the Noether charge,
and develop analytical mechanics for 
the system containing both bosonic and fermionic non-hermitian variables.
In Appendix B, we study the system of a bosonic spinor field
and clarify the difficulty on quantization.

\section{Harmonic oscillators}
\label{Harmonic oscillators}

First, we consider the systems of harmonic oscillators in quantum mechanics,
because they have features in common with those of scalar fields
and its study facilitates
the understanding of a difficulty and a remedy
appearing in a case with abnormal quantization rules
in quantum field theory.

\subsection{Ordinary harmonic oscillators}
\label{Ordinary harmonic oscillators}

For the sake of completeness,
let us begin with the system described by the Lagrangian,
\begin{eqnarray}
L_q = m \dot{q}^{\dagger} \dot{q} - m \omega^2 q^{\dagger} q~,
\label{L-q}
\end{eqnarray}
where $q=q(t)$ is a coordinate taking complex numbers, 
$q^{\dagger}$ its hermitian conjugate, $\dot{q} = dq/dt$,
and $\omega$ is an angular frequency.
The Euler-Lagrange equations for $q$ and $q^{\dagger}$ are given by
\begin{eqnarray}
m \frac{d^2 q^{\dagger}}{dt^2} =  - m \omega^2 q^{\dagger}~,~~
m \frac{d^2 q}{dt^2} =  - m \omega^2 q~,
\label{q-eq}
\end{eqnarray}
respectively.
They describe two harmonic oscillators with the same mass $m$.

According to Appendix A, let us define the canonical conjugate momenta of $q$ and $q^{\dagger}$ as
\begin{eqnarray}
p \equiv \left( \frac{\partial L_q}{\partial \dot{q}}\right)_{\rm R} = m \dot{q}^{\dagger}~,~~
p^{\dagger} \equiv 
\left(\frac{\partial L_q}{\partial \dot{q}^{\dagger}}\right)_{\rm L} = m \dot{q}~,
\label{p}
\end{eqnarray}
respectively.
Here, R and L stand for the right-differentiation and the left-differentiation, respectively.

By solving (\ref{q-eq}) and (\ref{p}), we obtain the solutions
\begin{eqnarray}
&~& q(t) = \sqrt{\frac{\hbar}{2 m \omega}} 
\left(a e^{-i\omega t} + b^{\dagger} e^{i\omega t}\right)~,~~
q^{\dagger}(t) = \sqrt{\frac{\hbar}{2 m \omega}} 
\left(a^{\dagger} e^{i\omega t} + b e^{-i\omega t}\right)~,
\label{q-dagger-sol}\\
&~& p(t) = i\sqrt{\frac{\hbar m\omega}{2}} 
\left(a^{\dagger} e^{i\omega t} - b e^{-i\omega t}\right)~,~~
p^{\dagger}(t) = - i\sqrt{\frac{\hbar m\omega}{2}} 
\left(a e^{-i\omega t} - b^{\dagger} e^{i\omega t}\right)~,
\label{p-dagger-sol}
\end{eqnarray}
where $a$, $b^{\dagger}$, $a^{\dagger}$ and $b$ are some constants.

Using (\ref{p}) and (\ref{H-def}), the Hamiltonian is obtained as
\begin{eqnarray}
H_q = p \dot{q} + \dot{q}^{\dagger} p^{\dagger} - L 
= \frac{1}{m} p p^{\dagger} + m \omega^2 q^{\dagger} q~.
\label{H-qp}
\end{eqnarray}
Later, we will comment on a consequence of a difference 
between $H_q$ and
the ordinary Hamiltonian defined by only the right-differentiation.

The system is quantized by regarding variables as operators
and imposing the following commutation relations 
on the canonical pairs $(q, p)$ and $(q^{\dagger}, p^{\dagger})$,
\begin{eqnarray}
&~& [q(t), p(t)] = i \hbar~,~~ [q^{\dagger}(t), p^{\dagger}(t)] = i \hbar~,~~
[q(t), p^{\dagger}(t)] = 0~,~~ 
\nonumber \\
&~& [q^{\dagger}(t), p(t)] = 0~,~~
[q(t), q^{\dagger}(t)] = 0~,~~ [p(t), p^{\dagger}(t)] = 0~,
\label{CCR-qp}
\end{eqnarray}
where $[A, B] \equiv AB-BA$.
Or equivalently, for operators
$a$, $b^{\dagger}$, $a^{\dagger}$ and $b$,
the following relations are imposed on,
\begin{eqnarray}
 [a, a^{\dagger}] =1~,~~ [b, b^{\dagger}] =1~,~~ [a, b] = 0~,~~ 
 [a^{\dagger}, b^{\dagger}] = 0~,~~
[a, b^{\dagger}] = 0~,~~ [a^{\dagger}, b] = 0~.
\label{CCR-ab}
\end{eqnarray}

In quantum theory, an operator $O=O(t)$ evolves by the Heisenberg equation,
\begin{eqnarray}
\frac{dO}{dt} = \frac{i}{\hbar} [H, O]~,
\label{H-eq}
\end{eqnarray}
where $H$ is the Hamiltonian.
From (\ref{H-qp}), (\ref{CCR-qp}) and (\ref{H-eq}), the following equations are derived
\begin{eqnarray}
&~& \frac{dq}{dt} = \frac{i}{\hbar} [H_q, q] = \frac{p^{\dagger}}{m}~,~~
\frac{dq^{\dagger}}{dt} = \frac{i}{\hbar} [H_q, q^{\dagger}] = \frac{p}{m}~,~~
\label{H-eq-qp1}\\
&~& \frac{dp}{dt} = \frac{i}{\hbar} [H_q, p] =- m \omega^2 q^{\dagger}~,~~
\frac{dp^{\dagger}}{dt} = \frac{i}{\hbar} [H_q, p^{\dagger}] =- m \omega^2 q~,
\label{H-eq-qp2}
\end{eqnarray}
where we use the relation among operators $A$, $B$ and $C$:
\begin{eqnarray}
[A B, C] = A[B, C] + [A, C]B~.
\label{R-C}
\end{eqnarray}
(\ref{H-eq-qp1}) and (\ref{H-eq-qp2}) are equivalent to (\ref{q-eq}) and (\ref{p}).

By inserting (\ref{q-dagger-sol}) and (\ref{p-dagger-sol}) into (\ref{H-qp}), $H_q$ is written by
\begin{eqnarray}
H_q = \hbar \omega \left(a^{\dagger} a + b b^{\dagger}\right)
= \hbar \omega \left(a^{\dagger} a + b^{\dagger} b + 1\right)~,
\label{H-ab}
\end{eqnarray}
where we use $[b, b^{\dagger}] =1$ to derive the last expression
and the constant part $\hbar \omega$ is the zero-point energy.
The eigenstates and eigenvalues of $H_q$ are given by
\begin{eqnarray}
\hspace{-0.2cm}
| n_a, n_b \rangle = \frac{(a^{\dagger})^{n_a}}{\sqrt{n_a !}}\frac{(b^{\dagger})^{n_b}}{\sqrt{n_b !}}
 |0, 0\rangle~;~
E_{n_a,n_b} =  \hbar \omega \left(n_a + n_b + 1\right)~,
\label{E-ab}
\end{eqnarray}
where $n_a, n_b = 0, 1, 2 \cdots$ and $|0, 0\rangle$ is the ground state
that satisfies $a |0, 0\rangle = 0$ and $b |0, 0\rangle = 0$.
We find that the values of $E_{n_a, n_b}$ are positive.
Using (\ref{CCR-ab}), the inner products are calculated as
\begin{eqnarray}
\langle m_a, m_b | n_a, n_b \rangle = \delta_{m_a n_a} \delta_{m_b n_b}~,
\label{norm}
\end{eqnarray}
and hence the positive-definiteness of norm holds on.
If we take $a^{\dagger} |0, 0\rangle = 0$ in place of $a |0, 0\rangle = 0$, 
$a |0, 0\rangle$ has a negative norm as seen from the relation,
\begin{eqnarray}
&~& 1 = \langle 0, 0|0, 0\rangle = 
\langle 0, 0| [a, a^{\dagger}] |0, 0\rangle 
= - \langle 0, 0| a^{\dagger} a |0, 0\rangle =- |a |0, 0\rangle |^2~.
\label{a-negative}
\end{eqnarray}
The same is true of $b$ and $b^{\dagger}$.

$L_q$ is invariant under the $U(1)$ transformation,
\begin{eqnarray}
\delta q = i [\epsilon N_q, q] = i \epsilon q~,~~
\delta q^{\dagger} = i [\epsilon N_q, q^{\dagger}]
= - i \epsilon q^{\dagger}~,
\label{U(1)}
\end{eqnarray}
where $\epsilon$ is an infinitesimal real number
and $N_q$ is the conserved $U(1)$ charge defined by
\begin{eqnarray}
\epsilon N_q 
\equiv \frac{1}{\hbar} \left[\left(\frac{\partial L_q}{\partial \dot{q}}\right)_{\rm R} \delta q
+ \delta q^{\dagger} \left(\frac{\partial L_q}{\partial \dot{q}^{\dagger}}\right)_{\rm L}\right]~.
\label{Q-def}
\end{eqnarray}
Note that $N_q$ is hermitian by definition, using (\ref{f-RL}) with $L_q^{\dagger} = L_q$.
From (\ref{Q-def}), $N_q$ is given by
\begin{eqnarray}
N_q = \frac{i}{\hbar} \left(p q- q^{\dagger} p^{\dagger}\right) = - a^{\dagger} a + b b^{\dagger}
=  - a^{\dagger} a + b^{\dagger} b + 1~,
\label{Q}
\end{eqnarray}
where we use $[b, b^{\dagger}] = 1$ to derive the last expression.

Here, we give a comment on a consequence 
of a difference between ours $(H_q, N_q)$ and
the following ordinary ones defined by only the right-differentiation,
\begin{eqnarray}
&~& H_{0} = p \dot{q} +  p^{\dagger} \dot{q}^{\dagger} - L_{q} 
= \frac{1}{m} p^{\dagger} p + m \omega^2 q^{\dagger} q~,~~
N_{0} = \frac{i}{\hbar} \left(pq- p^{\dagger}q^{\dagger}\right)~.
\label{N0}
\end{eqnarray}
Note that $N_{0}$ turns out to be hermitian
by using the commutation relations (\ref{CCR-qp}),
\begin{eqnarray}
N_{0}^{\dagger} = -\frac{i}{\hbar} \left(q^{\dagger} p^{\dagger}- q p\right)
= -\frac{i}{\hbar} \left(p^{\dagger} q^{\dagger}- p q\right) =N_{0}~.
\label{Q-dagger}
\end{eqnarray}
Using (\ref{q-dagger-sol}) and (\ref{p-dagger-sol}),
$H_{0}$ and $N_{0}$ are rewritten by
\begin{eqnarray}
\hspace{-1cm}&~& H_{0} 
= \frac{1}{2} \hbar \omega \left(a^{\dagger} a + a a^{\dagger} 
+ b^{\dagger} b + b b^{\dagger}\right)
= \hbar \omega \left(a^{\dagger} a + b^{\dagger} b + 1\right)~,
\label{H-ab-ordinary}\\
\hspace{-1cm}&~& N_{0} 
= \frac{1}{2}\left(- a^{\dagger} a - a a^{\dagger} + b^{\dagger} b + b b^{\dagger}\right)
= - a^{\dagger} a + b^{\dagger} b~,
\label{Q-ab-ordinary}
\end{eqnarray}
where we use $[a, b] = 0$ and $[a^{\dagger}, b^{\dagger}] = 0$ 
to derive the second expressions
and $[a, a^{\dagger}] =1$ and $[b, b^{\dagger}] =1$
to derive the last expressions.
In case that variables are bosonic or follow the commutation relations,
$H_0$ agrees with $H_q$ because of $p^{\dagger} p = p p^{\dagger}$
and we obtain the same result.
If we replace $[a, a^{\dagger}] =1$ and $[b, b^{\dagger}] =1$
with the anti-commutation relations 
$a a^{\dagger} + a^{\dagger} a=1$ and $b b^{\dagger} + b^{\dagger} b=1$,
we arrive at a misleading result 
that $H_0$ are $N_0$ are some constants from the second expressions
in (\ref{H-ab-ordinary}) and (\ref{Q-ab-ordinary}).
We will find that the treatment is not proper with careful consideration.

\subsection{Fermionic harmonic oscillators}
\label{Fermionic harmonic oscillators}

Next, we study the system described by the Lagrangian,
\begin{eqnarray}
L_{\xi} = m \dot{\xi}^{\dagger} \dot{\xi} - m \omega^2 \xi^{\dagger} \xi~,
\label{L-xi}
\end{eqnarray}
where $\xi=\xi(t)$ is a fermionic coordinate
taking Grassmann numbers and $L_{\xi}^{\dagger} = L_{\xi}$.
The Euler-Lagrange equations for $\xi$ and $\xi^{\dagger}$ are given by 
\begin{eqnarray}
m \frac{d^2 \xi^{\dagger}}{dt^2} =  - m \omega^2 \xi^{\dagger}~,~~
m \frac{d^2 \xi}{dt^2} =  - m \omega^2 \xi~,
\label{xi-eq}
\end{eqnarray}
respectively.
They also describe two harmonic oscillators with the same mass.

According to Appendix A, let us define
the canonical conjugate momenta of $\xi$ and $\xi^{\dagger}$ as
\begin{eqnarray}
\rho \equiv \left(\frac{\partial L_{\xi}}{\partial \dot{\xi}}\right)_{\rm R} 
= m \dot{\xi}^{\dagger}~,~~
\rho^{\dagger} \equiv
\left(\frac{\partial L_{\xi}}{\partial \dot{\xi}^{\dagger}}\right)_{\rm L} 
= m \dot{\xi}~,
\label{rho}
\end{eqnarray}
respectively.
By solving (\ref{xi-eq}) and (\ref{rho}), we obtain the solutions
\begin{eqnarray}
&~& \xi(t) = \sqrt{\frac{\hbar}{2 m \omega}} 
\left(c e^{-i\omega t} + d^{\dagger} e^{i\omega t}\right)~,~~
\xi^{\dagger}(t) = \sqrt{\frac{\hbar}{2 m \omega}} 
\left(c^{\dagger} e^{i\omega t} + d e^{-i\omega t}\right)~,
\label{xi-dagger-sol}\\
&~& \rho(t) = i\sqrt{\frac{\hbar m\omega}{2}} 
\left(c^{\dagger} e^{i\omega t} - d e^{-i\omega t}\right)~,~~
\rho^{\dagger}(t) = - i\sqrt{\frac{\hbar m\omega}{2}} 
\left(c e^{-i\omega t} - d^{\dagger} e^{i\omega t}\right)~,
\label{rho-dagger-sol}
\end{eqnarray}
where $c$, $d^{\dagger}$, $c^{\dagger}$ and $d$ are Grassmann numbers.

Using (\ref{rho}), the Hamiltonian is obtained as
\begin{eqnarray}
H_{\xi} = \rho \dot{\xi} + \dot{\xi}^{\dagger} \rho^{\dagger} - L_{\xi} 
= \frac{1}{m} \rho \rho^{\dagger} + m \omega^2 \xi^{\dagger} \xi~.
\label{H-xi}
\end{eqnarray}

Let us quantize the system by regarding variables as operators
and imposing the following anti-commutation relations 
on $(\xi, \rho)$ and $(\xi^{\dagger}, \rho^{\dagger})$,
\begin{eqnarray}
&~& \{\xi(t), \rho(t)\} = i \hbar~,~~ \{\xi^{\dagger}(t), \rho^{\dagger}(t)\} = -i \hbar~,~~
\{\xi(t), \xi(t)\} = 0~,~~ \{\rho(t), \rho(t)\} = 0~,~~ 
\nonumber \\
&~& \{\xi^{\dagger}(t), \xi^{\dagger}(t)\} = 0~,~~
\{\rho^{\dagger}(t), \rho^{\dagger}(t)\} = 0~,~~
\{\xi(t), \xi^{\dagger}(t)\} = 0~,~~
\nonumber \\
&~& \{\rho(t), \rho^{\dagger}(t)\} = 0~,~~
\{\xi(t), \rho^{\dagger}(t)\} = 0~,~~
\{\xi^{\dagger}(t), \rho(t)\} = 0~,
\label{CCR-xi}
\end{eqnarray}
where $\{A, B\} = AB+BA$.
Note that (\ref{CCR-xi}) are compatible with 
the classical counterparts (\ref{Poisson}) and (\ref{Quantization}).
Or equivalently, for operators $c$, $d^{\dagger}$, $c^{\dagger}$ and $d$,
the following relations are imposed on,
\begin{eqnarray}
&~& \{c, c^{\dagger}\} = 1~,~~ \{d, d^{\dagger}\} = -1~,~~ 
\{c, c\} = 0~,~~ 
\{c^{\dagger}, c^{\dagger}\} = 0~,~~
\{d, d\} = 0~,~~ \{d^{\dagger}, d^{\dagger}\} = 0~,~~ 
\nonumber \\
&~& \{c, d\} = 0~,~~ 
\{c^{\dagger}, d^{\dagger}\} = 0~,~~
\{c, d^{\dagger}\} = 0~,~~ \{c^{\dagger}, d\} =0~.
\label{CCR-cd}
\end{eqnarray}

From (\ref{H-xi}), (\ref{CCR-xi}) and (\ref{H-eq}), the following equations are derived
\begin{eqnarray}
&~& \frac{d\xi}{dt} = \frac{i}{\hbar} [H_{\xi}, \xi] = \frac{\rho^{\dagger}}{m}~,~~
\frac{d\xi^{\dagger}}{dt} = \frac{i}{\hbar} [H_{\xi}, \xi^{\dagger}] = \frac{\rho}{m}~,~~
\label{H-eq-xi1}\\
&~& 
\frac{d\rho}{dt} = \frac{i}{\hbar} [H_{\xi}, \rho] =- m \omega^2 \xi^{\dagger}~,~~
\frac{d\rho^{\dagger}}{dt} = \frac{i}{\hbar} [H_{\xi}, \rho^{\dagger}] =- m \omega^2 \xi~,
\label{H-eq-xi2}
\end{eqnarray}
where we use the relation,
\begin{eqnarray}
[A B, C] = A\{B, C\} - \{A, C\}B~.
\label{R-AC}
\end{eqnarray}
(\ref{H-eq-xi1}) and (\ref{H-eq-xi2}) are
equivalent to (\ref{xi-eq}) and (\ref{rho}).

By inserting (\ref{xi-dagger-sol}) and (\ref{rho-dagger-sol}) into (\ref{H-xi}), $H_{\xi}$ is written by
\begin{eqnarray}
H_{\xi} = \hbar \omega \left(c^{\dagger} c + d d^{\dagger}\right)
= \hbar \omega \left(c^{\dagger} c - d^{\dagger} d - 1\right)~,
\label{H-cd}
\end{eqnarray}
where we use $\{d, d^{\dagger}\} = -1$ to derive the last expression.

There are four eigenstates with the following eigenvalues for $H_{\xi}$
\begin{eqnarray}
&~& | 0, 0 \rangle~;~~ E=-\hbar \omega~,~~
\nonumber \\
&~& |1, 0 \rangle = c^{\dagger} | 0, 0 \rangle~,~~ |0, 1 \rangle = d^{\dagger} | 0, 0 \rangle~;~~ E=0~,~~
\nonumber \\
&~& |1, 1 \rangle = c^{\dagger} d^{\dagger} | 0, 0 \rangle~;~~ E=\hbar \omega~,
\label{E-cd}
\end{eqnarray}
where $|0, 0\rangle$ is the ground state
that satisfies $c |0, 0\rangle = 0$ and $d |0, 0\rangle = 0$.

From the relation,
\begin{eqnarray}
&~& 1 = \langle 0, 0|0, 0\rangle =
- \langle 0, 0| \{d, d^{\dagger}\} |0, 0\rangle
= - \langle 0, 0| d d^{\dagger} |0, 0\rangle
=- |d^{\dagger} |0, 0\rangle |^2~,
\label{negative}
\end{eqnarray}
we find that the state $d^{\dagger} |0, 0\rangle$ has a negative norm,
and then the probability interpretation does not hold on.
Hence, it is difficult to construct
a quantum theory for fermionic harmonic oscillators alone.

Even if we take another state $\widetilde{|0, 0 \rangle}$ as the ground state
that satisfies $c \widetilde{|0, 0\rangle} = 0$ and $d^{\dagger} \widetilde{|0, 0\rangle} = 0$,
the appearance of the negative norm states is inevitable from the relation,
\begin{eqnarray}
&~& 1 = \widetilde{\langle 0, 0}|\widetilde{0, 0\rangle}
=-\widetilde{\langle 0, 0|} \{d, d^{\dagger}\} \widetilde{|0, 0\rangle} 
= -\widetilde{\langle 0, 0|} d^{\dagger} d \widetilde{|0, 0\rangle}
= - |d \widetilde{|0, 0\rangle} |^2~,
\label{negative-tilde}
\end{eqnarray}
In this case, the energy spectrum is given by
\begin{eqnarray}
&~& d \widetilde{| 0, 0 \rangle}~;~~ E=-\hbar \omega~,~~
\nonumber \\
&~& \widetilde{|0, 0 \rangle} ~,~~  c^{\dagger} d \widetilde{| 0, 0 \rangle}~;~~ E=0~,~~
\nonumber \\
&~& c^{\dagger} \widetilde{| 0, 0 \rangle}~;~~ E=\hbar \omega~,
\label{E-cd-tilde}
\end{eqnarray}

$L_{\xi}$ is invariant under the $U(1)$ transformation,
\begin{eqnarray}
\delta \xi = i [\epsilon N_{\xi}, \xi] = i \epsilon \xi~,~~
\delta \xi^{\dagger} = i [\epsilon N_{\xi}, \xi^{\dagger}]
= - i \epsilon \xi^{\dagger}~,
\label{U(1)-xi}
\end{eqnarray}
where $N_{\xi}$ is the conserved $U(1)$ charge defined by
\begin{eqnarray}
\epsilon N_{\xi} 
\equiv \frac{1}{\hbar} \left[\left(\frac{\partial L_{\xi}}{\partial \dot{\xi}}\right)_{\rm R} \delta \xi
+ \delta \xi^{\dagger} \left(\frac{\partial L_{\xi}}{\partial \dot{\xi}^{\dagger}}\right)_{\rm L}\right]~.
\label{Q-xi-def}
\end{eqnarray}
From (\ref{Q-xi-def}), $N_{\xi}$ is given by
\begin{eqnarray}
N_{\xi} = \frac{i}{\hbar} \left(\rho \xi- \xi^{\dagger} \rho^{\dagger}\right) 
= - c^{\dagger} c + d d^{\dagger}
=  - c^{\dagger} c - d^{\dagger} d - 1~,
\label{Q-xi}
\end{eqnarray}
where we use $\{d, d^{\dagger}\}=-1$ to derive the last expression.

We find that both $H_{\xi}$ and $N_{\xi}$ are hermitian by definition 
and are not constants,
but the system is abnormal because it contains a state with a negative norm.
We point out that the same conclusion is obtained
even if we adopt the ordinary convention that the canonical momenta are
defined by only the right differentiation.

\subsection{Coexisting system of harmonic oscillators}
\label{Coexisting system of harmonic oscillators}

Next, let us consider the system that $(q, q^{\dagger})$ and $(\xi, \xi^{\dagger})$ coexist,
whose Lagrangian is given by
\begin{eqnarray}
L_{q, \xi} = m \dot{q}^{\dagger} \dot{q} - m \omega^2 q^{\dagger} q
+ m \dot{\xi}^{\dagger} \dot{\xi} - m \omega^2 \xi^{\dagger} \xi~.
\label{L-qxi}
\end{eqnarray}
From (\ref{H-ab}) and (\ref{H-cd}), the Hamiltonian is obtained as
\begin{eqnarray}
H_{q, \xi} = \hbar \omega \left(a^{\dagger} a + b^{\dagger} b + c^{\dagger} c - d^{\dagger} d\right)~.
\label{H-abcd}
\end{eqnarray} 
Note that the sum of the zero-point energies vanishes due to the cancellation between 
contributions from  $(q, q^{\dagger})$ and $(\xi, \xi^{\dagger})$.

There are four kinds of eigenstates with the following eigenvalues for $H_{q, \xi}$,
\begin{eqnarray}
&~& |n_a, n_b, 0, 0 \rangle 
\equiv \frac{(a^{\dagger})^{n_a}}{\sqrt{n_a !}}\frac{(b^{\dagger})^{n_b}}{\sqrt{n_b !}}
 |0, 0, 0, 0\rangle~;~~ 
E=\hbar \omega(n_a + n_b)~,~~
\label{E-1}\\
&~& |n_a, n_b, 1, 0 \rangle 
\equiv c^{\dagger} \frac{(a^{\dagger})^{n_a}}{\sqrt{n_a !}}\frac{(b^{\dagger})^{n_b}}{\sqrt{n_b !}}
 |0, 0, 0, 0\rangle~;~~ 
E=\hbar \omega(n_a + n_b +1)~,~~
\label{E-2}\\
&~& |n_a, n_b, 0, 1 \rangle 
\equiv d^{\dagger} \frac{(a^{\dagger})^{n_a}}{\sqrt{n_a !}}\frac{(b^{\dagger})^{n_b}}{\sqrt{n_b !}}
 |0, 0, 0, 0\rangle~;~~ 
E=\hbar \omega(n_a + n_b +1)~,~~
\label{E-3}\\
&~& |n_a, n_b, 1, 1 \rangle 
\equiv c^{\dagger} d^{\dagger} 
\frac{(a^{\dagger})^{n_a}}{\sqrt{n_a !}}\frac{(b^{\dagger})^{n_b}}{\sqrt{n_b !}}
 |0, 0, 0, 0\rangle~;~~ 
E=\hbar \omega(n_a + n_b +2)~,~~
\label{E-4}
\end{eqnarray}
where $|0, 0, 0, 0\rangle$ is the ground state
that satisfies $a |0, 0, 0, 0\rangle = 0$, $b |0, 0, 0, 0\rangle = 0$,
$c |0, 0, 0, 0\rangle = 0$ and $d |0, 0, 0, 0\rangle = 0$.

As seen from (\ref{negative}), $|n_a, n_b, 0, 1 \rangle$ and $|n_a, n_b, 1, 1 \rangle $
have a negative norm, and the theory appears to be inconsistent.
We will show that the system has fermionic symmetries 
and they save it from the disaster.

The $L_{q, \xi}$ is invariant under the fermionic transformations,
\begin{eqnarray}
\delta_{\rm F} q = -\zeta \xi~,~~\delta_{\rm F} q^{\dagger} = 0~,~~ 
\delta_{\rm F} \xi  = 0~,~~
\delta_{\rm F} \xi^{\dagger} = \zeta q^{\dagger}
\label{delta-F}
\end{eqnarray}
and 
\begin{eqnarray}
\delta_{\rm F}^{\dagger} q = 0~,~~
\delta_{\rm F}^{\dagger} q^{\dagger} = \zeta^{\dagger} \xi^{\dagger}~,~~
\delta_{\rm F}^{\dagger} \xi = \zeta^{\dagger} q~,~~
\delta_{\rm F}^{\dagger} \xi^{\dagger} = 0~,
\label{delta-Fdagger}
\end{eqnarray}
where $\zeta$ and $\zeta^{\dagger}$ are Grassmann numbers.
Note that $\delta_{\rm F}$ is not generated by a hermitian operator, 
different from the generator of the BRST transformation
in systems with first class constraints~\cite{BRST} 
and that of the topological symmetry~\cite{W,Top}.

From the above transformation properties, 
we see that ${\bm \delta}_{\rm F}$ and ${\bm \delta}_{\rm F}^{\dagger}$ are nilpotent, i.e.,
${\bm \delta}_{\rm F}^2= 0$ and ${\bm \delta}_{\rm F}^{\dagger2}=0$, or
\begin{eqnarray}
{Q_{\rm F}}^2 = 0~~~ {\rm and}~~~ {Q_{\rm F}^{\dagger}}^2 = 0~,
\label{QQ=0}
\end{eqnarray}
where the bold ones
${\bm \delta}_{\rm F}$ and ${\bm \delta}_{\rm F}^{\dagger}$ are 
defined by $\delta_{\rm F} = \zeta {\bm \delta}_{\rm F}$ 
and $\delta_{\rm F}^{\dagger} = \zeta^{\dagger} {\bm \delta}_{\rm F}^{\dagger}$, respectively.
$Q_{\rm F}$ and $Q_{\rm F}^{\dagger}$ are the corresponding generators given by
\begin{eqnarray}
\delta_{\rm F} O = i[\zeta Q_{\rm F}, O]~,~~
\delta_{\rm F}^{\dagger} O = i[Q_{\rm F}^{\dagger}\zeta^{\dagger}, O]~,
\label{delta-F-A}
\end{eqnarray}
and defined by
\begin{eqnarray}
\zeta Q_{\rm F} 
\equiv \frac{1}{\hbar} \left[\left(\frac{\partial L_{q, \xi}}{\partial \dot{q}}\right)_{\rm R} 
\delta_{\rm F} q
+ \delta_{\rm F} \xi^{\dagger} 
\left(\frac{\partial L_{q, \xi}}{\partial \dot{\xi}^{\dagger}}\right)_{\rm L}\right]~,~~
Q_{\rm F}^{\dagger} \zeta^{\dagger} 
\equiv \frac{1}{\hbar} \left[\delta_{\rm F}^{\dagger} q^{\dagger}
 \left(\frac{\partial L_{q, \xi}}{\partial \dot{q}^{\dagger}}\right)_{\rm L} 
+
\left(\frac{\partial L_{q, \xi}}{\partial \dot{\xi}}\right)_{\rm R}
\delta_{\rm F}^{\dagger} \xi \right]~.
\label{Q-F-dagger-def}
\end{eqnarray}

Furthermore, we find the algebraic relation,
\begin{eqnarray}
\{Q_{\rm F}, Q_{\rm F}^{\dagger}\} = N_{\rm D}~,~~
\label{QQdagger}
\end{eqnarray}
where $N_{\rm D}$ is the number operator
defined by
\begin{eqnarray}
N_{\rm D} \equiv -N_q - N_{\xi} 
= a^{\dagger} a - b^{\dagger} b  + c^{\dagger} c + d^{\dagger} d~.
\label{ND}
\end{eqnarray}
$N_q$ and $N_{\xi}$ are generators
for $U(1)$ transformations of $q$ and $\xi$, defined by
(\ref{Q}) and (\ref{Q-xi}), respectively.
The symmetry of our system is equivalent to $OSp(2|2)$.

From (\ref{Q-F-dagger-def}),
the conserved fermionic charges
$Q_{\rm F}$ and $Q_{\rm F}^{\dagger}$ are given by
\begin{eqnarray}
 Q_{\rm F} = \frac{1}{\hbar}\left( - p \xi + q^{\dagger} \rho^{\dagger}\right)~,~~
 Q_{\rm F}^{\dagger} = \frac{1}{\hbar}\left( - \xi^{\dagger} p^{\dagger} + \rho q\right)~.
\label{QF}
\end{eqnarray}
Then, the canonical momenta are transformed as,
\begin{eqnarray}
\delta_{\rm F} p = 0~,~~\delta_{\rm F} p^{\dagger} = -\zeta \rho^{\dagger}~,~~ 
\delta_{\rm F} \rho  = \zeta p~,~~
\delta_{\rm F} \rho^{\dagger} = 0~
\label{delta-F-p}
\end{eqnarray}
and 
\begin{eqnarray}
\delta_{\rm F}^{\dagger} p = \zeta^{\dagger} \rho~,~~
\delta_{\rm F}^{\dagger} p^{\dagger} = 0 ~,~~
\delta_{\rm F}^{\dagger} \rho = 0~,~~
\delta_{\rm F}^{\dagger} \rho^{\dagger} = -\zeta^{\dagger} p^{\dagger}~.
\label{delta-Fdagger-p}
\end{eqnarray}

It is easily understood that $L_{q, \xi}$ is invariant 
under the transformations (\ref{delta-F}) and (\ref{delta-Fdagger}),
from the nilpotency of ${\bm \delta}_{\rm F}$ and ${\bm \delta}_{\rm F}^{\dagger}$ 
and the relations,
\begin{eqnarray}
L_{q, \xi} =  {\bm \delta}_{\rm F}  {R}_{q, \xi}
= {\bm \delta}_{\rm F}^{\dagger} {R}_{q, \xi}^{\dagger}
= {\bm \delta}_{\rm F}  {\bm \delta}_{\rm F}^{\dagger} {L}_{q}
= - {\bm \delta}_{\rm F}^{\dagger}  {\bm \delta}_{\rm F} {L}_{q}~,
\label{L-qxi-exact}
\end{eqnarray}
where ${R}_{q, \xi}$ and ${R}_{q, \xi}^{\dagger}$ are given by
\begin{eqnarray}
{R}_{q, \xi} = m \dot{\xi}^{\dagger} \dot{q} - m \omega^2 \xi^{\dagger} q ~,~~
{R}_{q, \xi}^{\dagger} = m \dot{q}^{\dagger} \dot{\xi} - m \omega^2 q^{\dagger} \xi~.
\label{R-qxi}
\end{eqnarray}
The Hamiltonian $H_{q, \xi}$ is written in the $Q_{\rm F}$ and $Q_{\rm F}^{\dagger}$
exact forms such that 
\begin{eqnarray}
H_{q, \xi} =  i \left\{Q_{\rm F}, \tilde{R}_{q, \xi}\right\}
 = - i \left\{Q_{\rm F}^{\dagger}, \tilde{R}_{q, \xi}^{\dagger}\right\}
= \left\{Q_{\rm F}, \left\{Q_{\rm F}^{\dagger}, H_{q}\right\}\right\}
 = - \left\{Q_{\rm F}^{\dagger}, \left\{Q_{\rm F}, H_{q}\right\}\right\}~,
\label{H-qxi-exact}
\end{eqnarray}
where $\tilde{R}_{q, \xi}$ and $\tilde{R}_{q, \xi}^{\dagger}$ are given by
\begin{eqnarray}
\tilde{R}_{q, \xi} = \frac{1}{m} \rho p^{\dagger} + m \omega^2 \xi^{\dagger} q ~,~~
\tilde{R}_{q, \xi}^{\dagger} = \frac{1}{m} p \rho^{\dagger} + m \omega^2 q^{\dagger} \xi~.
\label{tildeR-qxi}
\end{eqnarray}

Using the solutions (\ref{q-dagger-sol}), (\ref{p-dagger-sol}), (\ref{xi-dagger-sol}),
and (\ref{rho-dagger-sol}),
$Q_{\rm F}$ and $Q_{\rm F}^{\dagger}$ are written by
\begin{eqnarray}
 Q_{\rm F} = - i (a^{\dagger} c - d^{\dagger} b)~,~~
 Q_{\rm F}^{\dagger} =  i (c^{\dagger} a - b^{\dagger} d)~.
\label{QF-abcd}
\end{eqnarray}

Then, the operators are transformed as,
\begin{eqnarray}
&~& \delta_{\rm F} a = -\zeta c~,~~\delta_{\rm F} a^{\dagger} = 0~,~~ 
\delta_{\rm F} b^{\dagger} = - d^{\dagger}~,~~
\nonumber \\
&~& \delta_{\rm F} c  = 0~,~~ \delta_{\rm F} c^{\dagger} = \zeta a^{\dagger}~,~~
\delta_{\rm F} d = \zeta b~,~~\delta_{\rm F} d^{\dagger} = 0
\label{delta-F-abcd}
\end{eqnarray}
and 
\begin{eqnarray}
&~& \delta_{\rm F}^{\dagger} a = 0~,~~
\delta_{\rm F}^{\dagger} a^{\dagger} = \zeta^{\dagger} c^{\dagger}~,~~
\delta_{\rm F}^{\dagger} b = \zeta^{\dagger} d~,~~
\delta_{\rm F}^{\dagger} b^{\dagger} = 0~,~~
\nonumber \\
&~& \delta_{\rm F}^{\dagger} c = \zeta^{\dagger} a~,~~
\delta_{\rm F}^{\dagger} c^{\dagger} = 0~,~~
\delta_{\rm F}^{\dagger} d = 0~,~~
\delta_{\rm F}^{\dagger} d^{\dagger} = \zeta^{\dagger} b^{\dagger}~.
\label{delta-Fdagger-abcd}
\end{eqnarray}

To formulate our model in a consistent manner,
we use a feature
that a conserved charge can, in general, be set to be zero as an auxiliary condition.
Let us select physical states $|{\rm phys}\rangle$
by imposing the following conditions on states,
\begin{eqnarray}
Q_{\rm F} |{\rm phys}\rangle = 0~,~~
Q_{\rm F}^{\dagger} |{\rm phys}\rangle = 0~,~~
N_{\rm D} |{\rm phys}\rangle = 0
\label{Phys}
\end{eqnarray}
or
\begin{eqnarray}
\tilde{Q}_1 |{\rm phys}\rangle = 0~,~~
\tilde{Q}_2 |{\rm phys}\rangle = 0~,~~
N_{\rm D} |{\rm phys}\rangle = 0~,
\label{Phys-tildeQ}
\end{eqnarray}
where $\tilde{Q}_1$ and $\tilde{Q}_2$ are hermitian fermionic charges 
defined by
\begin{eqnarray}
\tilde{Q}_1 \equiv Q_{\rm F} + Q_{\rm F}^{\dagger}~,~~
\tilde{Q}_2 \equiv i(Q_{\rm F} - Q_{\rm F}^{\dagger})~.
\label{Q12}
\end{eqnarray}
Then, the following relations are derived,
\begin{eqnarray}
&~& \langle {\rm phys} | \delta_{\rm F} O |{\rm phys} \rangle
 =\langle {\rm phys} | i[\zeta Q_{\rm F}, O] |{\rm phys} \rangle
= 0~,~~
\label{<delta-F-A>}\\
&~& \langle {\rm phys} | \delta_{\rm F}^{\dagger} O |{\rm phys} \rangle 
=\langle {\rm phys} | i[Q_{\rm F}^{\dagger}\zeta^{\dagger}, O] |{\rm phys} \rangle
= 0~.
\label{<delta-Fdagger-A>}
\end{eqnarray}
Using (\ref{<delta-F-A>}) and (\ref{<delta-Fdagger-A>}),
we obtain the following relations
from (\ref{delta-F-abcd}) and (\ref{delta-Fdagger-abcd}):
\begin{eqnarray}
&~& \langle {\rm phys}|a|{\rm phys}\rangle = 0~,~~
\langle {\rm phys}|a^{\dagger}|{\rm phys}\rangle = 0~,~~
\label{Phys-a}\\
&~& \langle {\rm phys}|b|{\rm phys}\rangle = 0~,~~
\langle {\rm phys}|b^{\dagger}|{\rm phys}\rangle = 0~,~~
\label{Phys-b}\\
&~& \langle {\rm phys}|c|{\rm phys}\rangle = 0~,~~
\langle {\rm phys}|c^{\dagger}|{\rm phys}\rangle = 0~,~~
\label{Phys-c}\\
&~& \langle {\rm phys}|d|{\rm phys}\rangle = 0~,~~
\langle {\rm phys}|d^{\dagger}|{\rm phys}\rangle = 0~.
\label{Phys-d}
\end{eqnarray}

The conditions (\ref{Phys}) are interpreted 
as counterparts of the Kugo-Ojima subsidiary condition
in the BRST quantization~\cite{K&O1,K&O2}.
We find that all states given by $|n_a, n_b, n_c, n_d \rangle$, 
except for the ground state $|0, 0, 0, 0 \rangle$, are
unphysical because they do not satisfy (\ref{Phys}).
Or it is interpreted as the quartet mechanism~\cite{K&O1,K&O2}.
The projection operator $P^{(n)}$ on the states with $n = n_a + n_b + n_c + n_d$
is given by
\begin{eqnarray}
&~& P^{(n)} = \frac{1}{n} \left(a^{\dagger} P^{(n-1)} a + b^{\dagger} P^{(n-1)} b \right.
\left. + c^{\dagger} P^{(n-1)} c - d^{\dagger} P^{(n-1)} d \right)~~~~(n \ge 1)~,
\label{P(n)}
\end{eqnarray}
and is written by
\begin{eqnarray}
P^{(n)} =  i \left\{Q_{\rm F}, R^{(n)}\right\}~,
\label{P(n)2}
\end{eqnarray}
where $R^{(n)}$ is given by
\begin{eqnarray}
R^{(n)} = \frac{1}{n} \left(c^{\dagger} P^{(n-1)} a + b^{\dagger} P^{(n-1)} d\right)~~~~(n \ge 1)~.
\label{R(n)}
\end{eqnarray}
We find that any state with $n \ge 1$ is unphysical from the relation
$\langle {\rm phys}|P^{(n)}|{\rm phys}\rangle = 0$ for  $n \ge 1$, i.e.,
only the ground state $|0, 0, 0, 0 \rangle$ is physical.
This is also regarded as a quantum mechanical version of the Parisi-Sourlas mechanism~\cite{P&S}.
The point is that the system has negative norm states, but they become unphysical
and harmless.

The remedy of the system described by $L_{\xi}$ is not unique.
There is a possibility that the real and the imaginary part
of $\xi$ are regarded as the Faddeev-Popov ghost variable $c(t)$ 
and the anti-ghost one $\overline{c}(t)$, respectively.
Using $\xi(t) = (c(t) + i \overline{c}(t))/\sqrt{2}$,
$L_{\xi}$ is rewritten by
\begin{eqnarray}
L_{\xi} =  m \dot{\xi}^{\dagger} \dot{\xi} - m \omega^2 \xi^{\dagger} \xi
= - i m \dot{\overline{c}} \dot{c} + i m \omega^2 \overline{c} c~.
\label{L-xi2}
\end{eqnarray}
We introduce those BRST partners $r(t)$ and $B(t)$
with the BRST transformation,
\begin{eqnarray}
\bm{\delta}_{\rm B} r = -c~,~~ \bm{\delta}_{\rm B} c = 0~,~~
\bm{\delta}_{\rm B} \overline{c} = i B~,~~ \bm{\delta}_{\rm B} B= 0~,
\label{BRST}
\end{eqnarray}
and construct the Lagrangian,
\begin{eqnarray}
&~& L_{r,c} = -i \bm{\delta}_{\rm B} \left(\dot{\overline{c}} \dot{r}
 - m\omega^2 \overline{c} r\right)
=  m \dot{B} \dot{r} - m \omega^2 B r
- i m \dot{\overline{c}} \dot{c} + i m \omega^2 \overline{c} c~.
\label{L-rc}
\end{eqnarray}
Using the change of variables $r(t) = (x(t) + y(t))/\sqrt{2}$
and $B(t) = (x(t) - y(t))/\sqrt{2}$,
the part containing $r(t)$ and $B(t)$ is rewritten as
\begin{eqnarray}
&~& L_{x,y} = m \dot{x}^2 - m \omega^2 x^2
- m \dot{y}^2 + m \omega^2 y^2~,
\label{L-xy}
\end{eqnarray}
and then we find that $x(t)$ has a positive norm and $y(t)$ has a negative norm.
Based on the BRST quantization, we understand that the system is also empty
leaving the vacuum state alone.

Furthermore, we give comments on similarities and differences between 
supersymmetric (SUSY) quantum mechanics\cite{Witten} and our model.
The ingredients of SUSY quantum mechanics 
are two hermitian fermionic charges $Q_i$
$(i=1, 2)$ that satisfy $Q_1 Q_2 + Q_2 Q_1 = 0$ 
and the Hamiltonian $H$ defined by $H= Q_1^2 = Q_2^2$.
By definition, $H$ is commutable to $Q_i$.
In our model, $\tilde{Q}_1$ and $\tilde{Q}_2$ satisfy 
$\tilde{Q}_1 \tilde{Q}_2 + \tilde{Q}_2 \tilde{Q}_1 = 0$
and $N_{\rm D} = \tilde{Q}_1^2 = \tilde{Q}_2^2$.
Here $N_{\rm D}$ is the number operator of the $Q_{\rm F}$ doublet.
Note that the algebraic relations among $(\tilde{Q}_1, \tilde{Q}_2, N_{\rm D})$
are same as $(Q_1, Q_2, H)$ in $N=2$ SUSY,
but $N_{\rm D}$ is different from our Hamiltonian $H_{q, \xi}$.

Our model is also formulated, using $\tilde{Q}_1$ and $\tilde{Q}_2$.
Concretely, $L_{q, \xi}$ and $H_{q, \xi}$ are written as
\begin{eqnarray}
&~& L_{q, \xi} = \bm{\delta}_1 R_{q, \xi}^{(1)} = \bm{\delta}_2 R_{q, \xi}^{(2)}~,~~
H_{q, \xi} = i \left\{\tilde{Q}_1, \tilde{R}_{q, \xi}^{(1)}\right\} 
=  i \left\{\tilde{Q}_2, \tilde{R}_{q, \xi}^{(2)}\right\}~,
\label{Hqxi-12}
\end{eqnarray}
where $\bm{\delta}_1$ and $\bm{\delta}_2$ are defined by
$\theta \bm{\delta}_1 O = i[\theta \tilde{Q}_1, O]
= i[\theta (Q_{\rm F} + Q_{\rm F}^{\dagger}), O]$
and $\theta \bm{\delta}_2 O = i[\theta \tilde{Q}_2, O]
= i[\theta i (Q_{\rm F} - Q_{\rm F}^{\dagger}), O]$
with a Grassmann parameter satisfying $\theta^{\dagger} = - \theta$.
$R_{q, \xi}^{(i)}$ and $\tilde{R}_{q, \xi}^{(i)}$
are given by
\begin{eqnarray}
&~& R_{q, \xi}^{(1)} = \frac{1}{2} (R_{q, \xi} - R_{q, \xi}^{\dagger})~,~~
R_{q, \xi}^{(2)} = \frac{1}{2i} (R_{q, \xi} + R_{q, \xi}^{\dagger})~,~~
\label{Rqxi-12}\\
&~& \tilde{R}_{q, \xi}^{(1)} = \frac{1}{2} (\tilde{R}_{q, \xi} - \tilde{R}_{q, \xi}^{\dagger})~,~~
\tilde{R}_{q, \xi}^{(2)} = \frac{1}{2i} (\tilde{R}_{q, \xi} + \tilde{R}_{q, \xi}^{\dagger})~.
\label{tildeRqxi-12}
\end{eqnarray}
They are anti-hermitian, i.e.,
$R_{q, \xi}^{(i)\dagger} = - R_{q, \xi}^{(i)}$ and $\tilde{R}_{q, \xi}^{(i)\dagger} = -\tilde{R}_{q, \xi}^{(i)}$,
and are invariant under the transformation generated by $N_{\rm D}$,
i. e., $\delta_{\rm D}R_{q, \xi}^{(i)} = 0$ and $[N_{\rm D}, \tilde{R}_{q, \xi}^{(i)}]=0$.
$\tilde{R}_{q, \xi}^{(1)}$ and $\tilde{R}_{q, \xi}^{(2)}$ satisfy the relations
$\tilde{R}_{q, \xi}^{(1) 2} =\tilde{R}_{q, \xi}^{(2) 2} =  - (\hbar \omega)^2 N_{\rm D}/4$ and 
$\tilde{R}_{q, \xi}^{(1)} \tilde{R}_{q, \xi}^{(2)} +
\tilde{R}_{q, \xi}^{(2)} \tilde{R}_{q, \xi}^{(1)} = 0$,
and $H_{q, \xi}$ is commutable to $\tilde{Q}_i$, $\tilde{R}_{q, \xi}^{(i)}$
and $N_{\rm D}$.

Every state has a positive norm in SUSY quantum mechanics 
and $H$ is positive semi-definite by definition.
In contrast, some states have a negative norm in our model
and hence $H_{q, \xi}$ is positive semi-definite despite its appearance.
The relation $N_{\rm D} = \tilde{Q}_1^2 = \tilde{Q}_2^2$
holds consistently with both positive and negative eigenvalues of $N_{\rm D}$,
because some states operated by $Q_i$ have negative norms.
In this way, our model is physically different from SUSY quantum mechanics.

\section{Scalar fields}
\label{Scalar fields}

\subsection{Ordinary scalar field}
\label{Ordinary scalar field}

Let us start with the system of a complex scalar field $\varphi$ described by the Lagrangian density,
\begin{eqnarray}
\mathcal{L}_{\varphi} = \partial_{\mu} \varphi^{\dagger} \partial^{\mu} \varphi 
- m^2 \varphi^{\dagger} \varphi~.
\label{L-varphi}
\end{eqnarray}
Here and hereafter, we use the metric tensor
$\eta_{\mu\nu}={\rm diag}(1,-1,-1,-1)$
and the natural units $c=\hbar=1$.
The Euler-Lagrange equations for $\varphi$ and $\varphi^{\dagger}$ are given by
\begin{eqnarray}
\left(\raisebox{-0.6mm}{\LBox} + m^2\right) \varphi^{\dagger} = 0~,~~ 
\left(\raisebox{-0.6mm}{\LBox} + m^2\right) \varphi = 0~,
\label{KG-eq}
\end{eqnarray}
and the canonical conjugate momenta of $\varphi$ and $\varphi^{\dagger}$ are defined by
\begin{eqnarray}
\pi \equiv \left(\frac{\partial \mathcal{L}}{\partial \dot{\varphi}}\right)_{\rm R}
=  \dot{\varphi}^{\dagger}~,~~
\pi^{\dagger} \equiv \left(\frac{\partial \mathcal{L}}{\partial \dot{\varphi}^{\dagger}}\right)_{\rm L}
= \dot{\varphi}~.
\label{pi}
\end{eqnarray}

By solving (\ref{KG-eq}) and (\ref{pi}), we obtain the solutions
\begin{eqnarray}
\hspace{-1cm}&~& \varphi(x) = \int \frac{d^3k}{\sqrt{(2\pi)^3 2k_0}}
\left(a(\bm{k}) e^{-i k x} + b^{\dagger}(\bm{k}) e^{i k x}\right)~,
\label{varphi-sol}\\
\hspace{-1cm}&~& \varphi^{\dagger}(x) = \int \frac{d^3k}{\sqrt{(2\pi)^3 2k_0}}
\left(a^{\dagger}(\bm{k}) e^{i k x} + b (\bm{k}) e^{-i k x}\right)~,
\label{varphi-dagger-sol}\\
\hspace{-1cm}&~& \pi(x) = i \int d^3k \sqrt{\frac{k_0}{2 (2\pi)^3}}
\left(a^{\dagger}(\bm{k}) e^{i k x} - b (\bm{k}) e^{-i k x}\right)~,
\label{pi-sol}\\
\hspace{-1cm}&~& \pi^{\dagger}(x) = - i \int d^3k \sqrt{\frac{k_0}{2 (2\pi)^3}}
\left(a(\bm{k}) e^{-i k x} - b^{\dagger} (\bm{k}) e^{i k x}\right)~,
\label{pi-dagger-sol}
\end{eqnarray}
where $k_0 = \sqrt{\bm{k}^2 + m^2}$ and $kx = k^{\mu} x_{\mu}$.

Using (\ref{pi}), the Hamiltonian density is obtained as
\begin{eqnarray}
\mathcal{H}_{\varphi} = \pi \dot{\varphi} 
+ \dot{\varphi}^{\dagger}\pi^{\dagger} - \mathcal{L}_{\varphi}
= \pi \pi^{\dagger} + \bm{\nabla} \varphi^{\dagger} \bm{\nabla} \varphi
+ m^2 \varphi^{\dagger} \varphi~.
\label{H-varphi}
\end{eqnarray}

The system is quantized by regarding variables as operators
and imposing the following commutation relations 
on the canonical pairs $(\varphi, \pi)$ and $(\varphi^{\dagger}, \pi^{\dagger})$,
\begin{eqnarray}
&~& [\varphi(\bm{x}, t), \pi(\bm{y}, t)] = i \delta^3(\bm{x}-\bm{y})~,~~
[\varphi^{\dagger}(\bm{x}, t), \pi^{\dagger}(\bm{y}, t)] = i \delta^3(\bm{x}-\bm{y})~,
\label{CCR-varphi}
\end{eqnarray}
and otherwise are zero.
Or equivalently, for operators $a(\bm{k})$, $b^{\dagger}(\bm{k})$, $a^{\dagger}(\bm{k})$ 
and $b(\bm{k})$,
the following commutation relations are imposed on,
\begin{eqnarray}
&~& [a(\bm{k}), a^{\dagger}(\bm{l})] = \delta^3(\bm{k}-\bm{l})~,~~ 
[b(\bm{k}), b^{\dagger}(\bm{l})] = \delta^3(\bm{k}-\bm{l})~,~~ 
[a(\bm{k}), b(\bm{l})] = 0~,~~ 
\nonumber \\
&~&[a^{\dagger}(\bm{k}), b^{\dagger}(\bm{l})] = 0~,~~
[a(\bm{k}), b^{\dagger}(\bm{l})] = 0~,~~ [a^{\dagger}(\bm{k}), b(\bm{l})] = 0~,~~
[a(\bm{k}), a(\bm{l})] = 0~,~~ 
\nonumber \\
&~& [a^{\dagger}(\bm{k}), a^{\dagger}(\bm{l})] = 0~,~~
[b(\bm{k}), b(\bm{l})] = 0~,~~ [b^{\dagger}(\bm{k}), b^{\dagger}(\bm{l})] = 0~.
\label{CCR-ab-varphi}
\end{eqnarray}

Using (\ref{R-C}), (\ref{H-varphi}), (\ref{CCR-varphi}) and the Heisenberg equation,
(\ref{KG-eq}) are derived where 
the Hamiltonian is given by $H_{\varphi} = \int \mathcal{H}_{\varphi} d^3x$.

By inserting (\ref{varphi-sol}) -- (\ref{pi-dagger-sol}) into (\ref{H-varphi}), 
the Hamiltonian $H_{\varphi}$ is written by
\begin{eqnarray}
\hspace{-1.3cm}&~& H_{\varphi} = \int d^3k k_0 
\left(a^{\dagger}(\bm{k}) a(\bm{k}) + b(\bm{k}) b^{\dagger}(\bm{k})\right)
\nonumber \\
\hspace{-1.3cm}&~& ~~~~~~~ = \int d^3k k_0 
\left(a^{\dagger}(\bm{k}) a(\bm{k}) + b^{\dagger}(\bm{k}) b(\bm{k})\right)
+ \int \frac{d^3k d^3x}{(2\pi)^3} k^0~.
\label{H-ab-varphi}
\end{eqnarray}
The ground state $| 0 \rangle$ is defined as the state that 
satisfies $a(\bm{k}) |0 \rangle = 0$ and $b(\bm{k}) |0 \rangle = 0$.
The eigenstates and eigenvalues of $H_{\varphi}$ are given by
\begin{eqnarray}
\hspace{-0.6cm}&~& \int d^3k_1 d^3k_2 \cdots d^3k_{n_a} d^3l_1 d^3l_2 \cdots d^3l_{n_b}
f_1(\bm{k}_1) f_2(\bm{k}_2) \cdots f_{n_a}(\bm{k}_{n_a}) 
g_1(\bm{l}_1) g_2(\bm{l}_2) \cdots g_{n_b}(\bm{l}_{n_b}) 
\nonumber \\
\hspace{-0.6cm}&~& ~~
\cdot a^{\dagger}(\bm{k}_1) a^{\dagger}(\bm{k}_2) \cdots a^{\dagger}(\bm{k}_{n_a})
b^\dagger(\bm{l}_1) b^{\dagger}(\bm{l}_2) \cdots b^{\dagger}(\bm{l}_{n_b}) |0\rangle~,~~
\label{H-varphi-states}\\
\hspace{-0.6cm}&~& E= k_{10} + k_{20} + \cdots + k_{n_a 0} + l_{10} + l_{20} + \cdots + l_{n_b 0}~,
\label{E-varphi}
\end{eqnarray}
where $f_n(\bm{k}_n)$ and $g_n(\bm{l}_n)$ are some square integrable functions,
$k_{n 0} = \sqrt{\bm{k}^2_n + m^2}$, $l_{n 0} = \sqrt{\bm{l}^2_n + m^2}$, and
we subtract an infinite constant corresponding to the sum of the zero-point energies
because not the energy itself
but the energy difference has physical meaning in the absence of gravity.
Concretely, using the normal ordering, we define $H_{\varphi}$ by
\begin{eqnarray}
H_{\varphi}  \equiv  : H_{\varphi} :  =
 \int d^3k k_0 \left(a^{\dagger}(\bm{k}) a(\bm{k}) + b^{\dagger}(\bm{k}) b(\bm{k})\right)~.
\label{H-varphi-normal}
\end{eqnarray}

$\mathcal{L}_{\varphi}$ is invariant under the $U(1)$ transformation,
\begin{eqnarray}
\delta \varphi = i [\epsilon N_{\varphi}, \varphi] = i \epsilon \varphi~,~~
\delta \varphi^{\dagger} = i [\epsilon N_{\varphi}, \varphi^{\dagger}]
= - i \epsilon \varphi^{\dagger}~,
\label{U(1)-varphi}
\end{eqnarray}
where $N_{\varphi}$ is the conserved $U(1)$ charge defined by
\begin{eqnarray}
\epsilon N_{\varphi} 
\equiv \int d^3x \left[
\left(\frac{\partial \mathcal{L}_{\varphi}}{\partial \dot{\varphi}}\right)_{\rm R} \delta \varphi
+ \delta \varphi^{\dagger} 
\left(\frac{\partial \mathcal{L}_{\varphi}}{\partial \dot{\varphi}^{\dagger}}\right)_{\rm L}
\right]~.
\label{Q-def-varphi}
\end{eqnarray}
Note that $N_{\varphi}$ is hermitian by definition and 
$\mathcal{L}_{\varphi}^{\dagger} = \mathcal{L}_{\varphi}$.
From (\ref{Q-def-varphi}), $N_{\varphi}$ is given by
\begin{eqnarray}
\hspace{-1.2cm}&~& N_{\varphi} 
= i \int d^3x \left(\pi \varphi - \varphi^{\dagger} \pi^{\dagger}\right) 
=\int d^3k \left(- a^{\dagger}(\bm{k}) a(\bm{k}) 
+ b(\bm{k}) b^{\dagger}(\bm{k})\right)
\nonumber \\
\hspace{-1.2cm}&~& ~~~~~~ = \int d^3k \left( - a^{\dagger}(\bm{k}) a(\bm{k}) 
+ b^{\dagger}(\bm{k}) b(\bm{k})\right) 
+ \int\frac{d^3k d^3x}{(2\pi)^3}~,
\label{Q-varphi}
\end{eqnarray}
where we use $[b(\bm{k}), b^{\dagger}(\bm{l})] = \delta^3(\bm{k}-\bm{l})$ 
to derive the last expression.
To subtract the infinite constant in $N_{\varphi}$, we define $N_{\varphi}$ by
\begin{eqnarray}
N_{\varphi}  \equiv  : N_{\varphi} :  =
 \int d^3k \left( - a^{\dagger}(\bm{k}) a(\bm{k}) + b^{\dagger}(\bm{k}) b(\bm{k})\right)~.
\label{Q-varphi-normal}
\end{eqnarray}
We find that the $U(1)$ charge of particle corresponding $b^{\dagger}(\bm{k}) | 0 \rangle$
is opposite to that corresponding $a^{\dagger}(\bm{k}) | 0 \rangle$.
Hence, $a(\bm{k})$ and $b^{\dagger}(\bm{k})$ are regarded as
the annihilation operator of particle and the creation operator of antiparticle, respectively.

The 4-dimensional commutation relations are calculated as
\begin{eqnarray}
&~& [\varphi(x), \varphi^{\dagger}(y)] 
= \int \frac{d^3k d^3l}{(2\pi)^3 \sqrt{2k_0~2l_0}} \left([a(\bm{k}), a^{\dagger}(\bm{l})]e^{-ikx+ily}
+ [b^{\dagger}(\bm{k}), b(\bm{l})]e^{ikx-ily}\right)
\nonumber \\
&~& ~~~~~~~~~~~~~~~~~~~~~~~~= \int \frac{d^3k d^3l}{(2\pi)^3 \sqrt{2k_0~2l_0}} 
\left([a(\bm{k}), a^{\dagger}(\bm{l})]e^{-ikx+ily}- [b(\bm{l}), b^{\dagger}(\bm{k})]e^{ikx-ily}\right)
\nonumber \\
&~& ~~~~~~~~~~~~~~~~~~~~~~~~  = \int \frac{d^3k}{(2\pi)^3 2k_0} \left(e^{-ik(x-y)} - e^{ik(x-y)}\right)
\nonumber \\
&~& ~~~~~~~~~~~~~~~~~~~~~~~~ 
= \int \frac{d^4k}{(2\pi)^3} \epsilon(k_0) \delta(k^2-m^2)
e^{-ik(x-y)} 
\equiv i \varDelta(x-y)~,~~ 
\label{Delta}\\
&~& [\varphi(x), \varphi(y)] = 0~,~~
[\varphi^{\dagger}(x), \varphi^{\dagger}(y)] = 0~,
\label{4D-CCR-varphi}
\end{eqnarray}
where $\epsilon(k_0) = k_0/|k_0|$ with $\epsilon(0)=0$,
$\varDelta(x-y)$ is the invariant delta function, and
two fields separated by a space-like interval commute with each other
as seen from the relation $\varDelta(x-y) = 0$ for $(x-y)^2 < 0$.
This feature is called $\lq$the microscopic causality'.

The vacuum expectation values of the time ordered products are calculated as
\begin{eqnarray}
&~& \langle 0 |{\rm T}\varphi(x) \varphi^{\dagger}(y)| 0 \rangle 
= \langle 0 |(\theta(x_0 - y_0) \varphi(x) \varphi^{\dagger}(y)
 + \theta(y_0 - x_0) \varphi^{\dagger}(y) \varphi(x))| 0 \rangle 
\nonumber \\
&~& ~~~~~~~~~~~~~~~~~~~~~~~~~~~~~~~~~ = \int \frac{d^3k d^3l}{(2\pi)^3 \sqrt{2k_0~2l_0}} 
\left(\theta(x_0- y_0)\langle 0 |a(\bm{k}) a^{\dagger}(\bm{l})| 0 \rangle e^{-ikx+ily} \right.
\nonumber \\
&~& ~~~~~~~~~~~~~~~~~~~~~~~~~~~~~~~~~~~~~~~~~~~~~~~~~~~~~~~~~~~~~~~~~~~~~~~~~~ 
\left. + \theta(y_0 - x_0) \langle 0 | b(\bm{k}) b^{\dagger}(\bm{l})| 0 \rangle e^{ikx-ily}\right)
\nonumber \\
&~& ~~~~~~~~~~~~~~~~~~~~~~~~~~~~~~~~~ = \int \frac{d^3k}{(2\pi)^3 2k_0} 
\left(\theta(x_0- y_0)e^{-ik(x-y)} + \theta(y_0- x_0) e^{ik(x-y)}\right)
\nonumber \\
&~& ~~~~~~~~~~~~~~~~~~~~~~~~~~~~~~~~~ 
= \int \frac{d^4k}{(2\pi)^4} \frac{i e^{-ik(x-y)}}{k^2 - m^2 + i \varepsilon}
\equiv i \varDelta_{\rm F}(x-y)~,~~ 
\label{Delta-F}\\
&~& \langle 0 |{\rm T}\varphi(x) \varphi(y)| 0 \rangle = 0~,~~
\langle 0 |{\rm T}\varphi^{\dagger}(x) \varphi^{\dagger}(y)| 0 \rangle = 0~,
\label{Tproduct-varphi}
\end{eqnarray}
where $\varDelta_{\rm F}(x-y)$ is the Feynman propagator.

Here, we roughly estimate what happens for the causality in the case with abnormal relations.
Using (\ref{varphi-sol}) and (\ref{varphi-dagger-sol}),
the 4-dimensional anti-commutation relation between $\varphi(x)$ and $\varphi^{\dagger}(y)$
is given by
\begin{eqnarray}
\hspace{-1cm} &~& \{\varphi(x), \varphi^{\dagger}(y)\} 
= \int \frac{d^3k d^3l}{(2\pi)^3 \sqrt{2k_0~2l_0}} \left(\{a(\bm{k}), a^{\dagger}(\bm{l})\}e^{-ikx+ily}
+ \{b^{\dagger}(\bm{k}), b(\bm{l})\}e^{ikx-ily}\right)~,
\nonumber \\
\hspace{-1cm} &~& ~~~~~~~~~~~~~~~~~~~~~~~~~ 
= \int \frac{d^3k d^3l}{(2\pi)^3 \sqrt{2k_0~2l_0}} \left(\{a(\bm{k}), a^{\dagger}(\bm{l})\}e^{-ikx+ily}
+ \{b(\bm{l}), b^{\dagger}(\bm{k})\}e^{ikx-ily}\right)~.
\label{4D-ACR-varphi}
\end{eqnarray}
If we replace
$[a(\bm{k}), a^{\dagger}(\bm{l})] = \delta^3(\bm{k} - \bm{l})$
and $[b(\bm{l}), b^{\dagger}(\bm{k})] = \delta^3(\bm{k} - \bm{l})$
with $\{a(\bm{k}), a^{\dagger}(\bm{l})\} = \delta^3(\bm{k} - \bm{l})$
and $\{b(\bm{l}), b^{\dagger}(\bm{k})\} = \delta^3(\bm{k} - \bm{l})$,
we obtain the relation,
\begin{eqnarray}
\hspace{-1cm}&~& \{\varphi(x), \varphi^{\dagger}(y)\} 
= \int \frac{d^3k}{(2\pi)^3 2k_0} \left(e^{-ik(x-y)} + e^{ik(x-y)}\right) 
\equiv i \varDelta^{(1)}(x-y)~.
\label{varDelta(1)}
\end{eqnarray}
(\ref{varDelta(1)}) means that the causality is violated
because $\varDelta^{(1)}(x-y)$ does not vanish for $(x-y)^2 < 0$.
However, it would still be unwise to conclude that
the causality is conflict with
the anti-commutation relations imposed on a complex scalar field,
because it is not clear whether the above replacement of relations is appropriate 
or compatible with relations including the field equations.
We will study it soon.

\subsection{Fermionic scalar field}
\label{Fermionic scalar field}

Let us study the system described by the Lagrangian density,
\begin{eqnarray}
\mathcal{L}_{c_{\varphi}} = \partial_{\mu} c_{\varphi}^{\dagger} \partial^{\mu} c_{\varphi} 
- m^2 c_{\varphi}^{\dagger} c_{\varphi}~,
\label{L-c}
\end{eqnarray}
where $c_{\varphi} = c_{\varphi}(x)$ is a complex scalar field taking Grassmann numbers. 
The Euler-Lagrange equations for $c_{\varphi}$ and $c_{\varphi}^{\dagger}$ 
are given by
\begin{eqnarray}
\left(\raisebox{-0.6mm}{\LBox} + m^2\right) c_{\varphi}^{\dagger} = 0~,~~ 
\left(\raisebox{-0.6mm}{\LBox} + m^2\right) c_{\varphi} = 0~,
\label{KG-eq-c}
\end{eqnarray}
respectively.

The canonical conjugate momenta of $c_{\varphi}$ and $c_{\varphi}^{\dagger}$ are defined by
\begin{eqnarray}
\pi_{c_{\varphi}} \equiv 
\left(\frac{\partial \mathcal{L}_{c_{\varphi}}}{\partial \dot{c}_{\varphi}}\right)_{\rm R}
=  \dot{c}_{\varphi}^{\dagger}~,~~
\pi_{c_{\varphi}}^{\dagger} 
\equiv \left(\frac{\partial \mathcal{L}_{c_{\varphi}}}{\partial \dot{c}_{\varphi}^{\dagger}}\right)_{\rm L}
= \dot{c}_{\varphi}~,
\label{pi-c}
\end{eqnarray}
respectively.

By solving (\ref{KG-eq-c}) and (\ref{pi-c}), we obtain the solutions
\begin{eqnarray}
\hspace{-1cm}&~& c_{\varphi}(x) = \int \frac{d^3k}{\sqrt{(2\pi)^3 2k_0}}
\left(c(\bm{k}) e^{-i k x} + d^{\dagger}(\bm{k}) e^{i k x}\right)~,
\label{c-sol}\\
\hspace{-1cm}&~& c_{\varphi}^{\dagger}(x) = \int \frac{d^3k}{\sqrt{(2\pi)^3 2k_0}}
\left(c^{\dagger}(\bm{k}) e^{i k x} + d (\bm{k}) e^{-i k x}\right)~,
\label{c-dagger-sol}\\
\hspace{-1cm}&~& \pi_{c_{\varphi}}(x) = i \int d^3k \sqrt{\frac{k_0}{2 (2\pi)^3}}
\left(c^{\dagger}(\bm{k}) e^{i k x} - d (\bm{k}) e^{-i k x}\right)~,
\label{pi-c-sol}\\
\hspace{-1cm}&~& \pi_{c_{\varphi}}^{\dagger}(x) = - i \int d^3k \sqrt{\frac{k_0}{2 (2\pi)^3}}
\left(c(\bm{k}) e^{-i k x} - d^{\dagger} (\bm{k}) e^{i k x}\right)~.
\label{pi-c-dagger-sol}
\end{eqnarray}

Using (\ref{pi-c}), the Hamiltonian density is obtained as
\begin{eqnarray}
&~& \mathcal{H}_{c_{\varphi}} = \pi_{c_{\varphi}} \dot{c}_{\varphi} 
+ \dot{c}_{\varphi}^{\dagger} \pi_{c_{\varphi}}^{\dagger}
- \mathcal{L}_{c_{\varphi}}
= \pi_{c_{\varphi}} \pi_{c_{\varphi}}^{\dagger} 
+ \bm{\nabla} c_{\varphi}^{\dagger} \bm{\nabla} c_{\varphi}
+ m^2 c_{\varphi}^{\dagger} c_{\varphi}~.
\label{H-c}
\end{eqnarray}

Let us quantize the system by regarding variables as operators
and imposing the following anti-commutation relations 
on $(c_{\varphi}, \pi_{c_{\varphi}})$ 
and $(c_{\varphi}^{\dagger}, \pi_{c_{\varphi}}^{\dagger})$,
\begin{eqnarray}
&~& \{c_{\varphi}(\bm{x}, t), \pi_{c_{\varphi}}(\bm{y}, t)\} = i \delta^3(\bm{x}-\bm{y})~,~~
\{c_{\varphi}^{\dagger}(\bm{x}, t), \pi_{c_{\varphi}}^{\dagger}(\bm{y}, t)\} = -i \delta^3(\bm{x}-\bm{y})~,
\label{CCR-c}
\end{eqnarray}
and otherwise are zero.
Or equivalently, 
for operators $c(\bm{k})$, $d^{\dagger}(\bm{k})$, $c^{\dagger}(\bm{k})$ and $d(\bm{k})$,
the following relations are imposed on,
\begin{eqnarray}
\hspace{-0.6cm}&~& \{c(\bm{k}), c^{\dagger}(\bm{l})\} = \delta^3(\bm{k}-\bm{l})~,~~
\{d(\bm{k}), d^{\dagger}(\bm{l})\} = - \delta^3(\bm{k}-\bm{l})~,~~ 
\{c(\bm{k}), c(\bm{l})\} = 0~,~~ 
\nonumber \\
\hspace{-0.6cm}&~&\{c^{\dagger}(\bm{k}), c^{\dagger}(\bm{l})\} = 0~,~~
\{d(\bm{k}), d(\bm{l})\} = 0~,~~ 
\{d^{\dagger}(\bm{k}), d^{\dagger}(\bm{l})\} = 0~,~~
\{c(\bm{k}), d(\bm{l})\} = 0~,~~ 
\nonumber \\
\hspace{-0.6cm}&~&
\{c^{\dagger}(\bm{k}), d^{\dagger}(\bm{l})\} = 0~,~~
\{c(\bm{k}), d^{\dagger}(\bm{l})\} = 0~,~~ \{c^{\dagger}(\bm{k}), d(\bm{l})\} = 0~.
\label{CCR-cd-c}
\end{eqnarray}

Using (\ref{R-AC}), (\ref{H-c}), (\ref{CCR-c}) and the Heisenberg equation, 
(\ref{KG-eq-c}) and (\ref{pi-c}) are derived
where the Hamiltonian is given by $H_{c_{\varphi}} = \int \mathcal{H}_{c_{\varphi}} d^3x$

By inserting (\ref{c-sol}) -- (\ref{pi-c-dagger-sol}) into (\ref{H-c}), 
the Hamiltonian $H_{c_{\varphi}}$ is written by
\begin{eqnarray}
\hspace{-1.3cm}&~& H_{c_{\varphi}} = \int d^3k k_0 
\left(c^{\dagger}(\bm{k}) c(\bm{k}) + d(\bm{k}) d^{\dagger}(\bm{k}) \right)
\nonumber \\
\hspace{-1.3cm}&~& ~~~~~~~~ = \int d^3k k_0 
\left(c^{\dagger}(\bm{k}) c(\bm{k}) - d^{\dagger}(\bm{k}) d(\bm{k})\right)
- \int \frac{d^3k d^3x}{(2\pi)^3} k^0~.
\label{H-cd-c}
\end{eqnarray}
The ground state $| 0 \rangle$ is defined by the state that 
satisfies $c(\bm{k}) |0 \rangle = 0$ and $d(\bm{k}) |0 \rangle = 0$.
The eigenstates and eigenvalues of $H_{c_{\varphi}}$ are given by
\begin{eqnarray}
\hspace{-0.7cm}&~& \int d^3k_1 d^3k_2 \cdots d^3k_{n_a} d^3l_1 d^3l_2 \cdots d^3l_{n_b}
f_1(\bm{k}_1) f_2(\bm{k}_2) \cdots f_{n_a}(\bm{k}_{n_a}) 
 g_1(\bm{l}_1) g_2(\bm{l}_2) \cdots g_{n_b}(\bm{l}_{n_b}) 
\nonumber \\
\hspace{-0.7cm}&~& ~~ \cdot
c^{\dagger}(\bm{k}_1) c^{\dagger}(\bm{k}_2) \cdots c^{\dagger}(\bm{k}_{n_a})
d^\dagger(\bm{l}_1) d^{\dagger}(\bm{l}_2) \cdots d^{\dagger}(\bm{l}_{n_b}) |0\rangle~,~~
\label{H-c-states}\\
\hspace{-0.7cm}&~& E= k_{10} + k_{20} + \cdots + k_{n_a 0} + l_{10} + l_{20} + \cdots + l_{n_b 0}~,
\label{E-c}
\end{eqnarray}
where we subtract an infinite constant corresponding to the sum of the zero-point energies.
We find that the energy is positive, although the anti-commutation relations are imposed on
scalar fields
and the negative sign appears in front of $d^{\dagger}(\bm{k}) d(\bm{k})$
in $H_{c_{\varphi}}$.
Note that the negative sign also exists in
$\{d(\bm{k}), d^{\dagger}(\bm{l})\} = - \delta^3(\bm{k}-\bm{l})$
and it guarantees the positivity of energy.

$\mathcal{L}_{c_{\varphi}}$ is invariant under the $U(1)$ transformation,
\begin{eqnarray}
\delta c_{\varphi} = i [\epsilon N_{c_{\varphi}}, c_{\varphi}] = i \epsilon c_{\varphi}~,~~
\delta c_{\varphi}^{\dagger} = i [\epsilon N_{c_{\varphi}}, c_{\varphi}^{\dagger}]
= - i \epsilon c_{\varphi}^{\dagger}~,
\label{U(1)-cvarphi}
\end{eqnarray}
where $N_{c_{\varphi}}$ is the conserved $U(1)$ charge defined by
\begin{eqnarray}
\epsilon N_{c_{\varphi}} 
\equiv \int d^3x \left[
\left(\frac{\partial \mathcal{L}_{c_{\varphi}}}{\partial \dot{c}_{\varphi}}\right)_{\rm R} 
\delta c_{\varphi}
+ \delta c_{\varphi}^{\dagger} 
\left(\frac{\partial \mathcal{L}_{c_{\varphi}}}{\partial \dot{c}_{\varphi}^{\dagger}}\right)_{\rm L}
\right]~.
\label{Q-def-cvarphi}
\end{eqnarray}
Note that $N_{c_{\varphi}}$ is hermitian by definition and 
$\mathcal{L}_{c_{\varphi}}^{\dagger} = \mathcal{L}_{c_{\varphi}}$.
From (\ref{Q-def-cvarphi}), $N_{c_{\varphi}}$ is given by
\begin{eqnarray}
\hspace{-1.2cm}&~& N_{c_{\varphi}} 
= i \int d^3x \left(\pi_{c_{\varphi}} c_{\varphi} 
- c_{\varphi}^{\dagger} \pi_{c_{\varphi}}^{\dagger}\right) 
=\int d^3k \left(-c^{\dagger}(\bm{k}) c(\bm{k}) 
+ d(\bm{k}) d^{\dagger}(\bm{k})\right)
\nonumber \\
\hspace{-1.2cm}&~& ~~~~~~~~ = - \int d^3k \left(c^{\dagger}(\bm{k}) c(\bm{k}) 
+ d^{\dagger}(\bm{k}) d(\bm{k})\right) 
- \int\frac{d^3k d^3x}{(2\pi)^3}~,
\label{Q-cvarphi}
\end{eqnarray}
where we use $\{d(\bm{k}), d^{\dagger}(\bm{l})\} = -\delta^3(\bm{k}-\bm{l})$ 
to derive the last expression.
To subtract the infinite constant in $N_{c_{\varphi}}$,
we define $N_{c_{\varphi}}$ by
\begin{eqnarray}
N_{c_{\varphi}} \equiv : N_{c_{\varphi}} : =
- \int d^3k \left(c^{\dagger}(\bm{k}) c(\bm{k}) + d^{\dagger}(\bm{k}) d(\bm{k})\right)~.
\label{Q-cvarphi-normal}
\end{eqnarray}

The 4-dimensional anti-commutation relations are calculated as
\begin{eqnarray}
&~& \{c_{\varphi}(x), c_{\varphi}^{\dagger}(y)\}
= \int \frac{d^3k d^3l}{(2\pi)^3 \sqrt{2k_0~2l_0}} \left(\{c(\bm{k}), c^{\dagger}(\bm{l})\}e^{-ikx+ily}
+ \{d^{\dagger}(\bm{k}), d(\bm{l})\} e^{ikx-ily}\right)
\nonumber \\
&~& ~~~~~~~~~~~~~~~~~~~~~~~~~~ = \int \frac{d^3k}{(2\pi)^3 2k_0} \left(e^{-ik(x-y)} - e^{ik(x-y)}\right)
= i \varDelta(x-y)~,~~ 
\label{Delta-c}\\
&~& \{c_{\varphi}(x), c_{\varphi}(y)\} = 0~,~~
\{c_{\varphi}^{\dagger}(x), c_{\varphi}^{\dagger}(y)\} = 0~,
\label{4D-CCR-c}
\end{eqnarray}
where we use the anti-commutation relations (\ref{CCR-cd-c}).
Then, bosonic variables composed of 
$c_{\varphi}$ and $c_{\varphi}^{\dagger}$ are commutative to any bosonic variables
separated by a space-like interval, and hence
the microscopic causality is not violated.
From (\ref{Delta}) and (\ref{Delta-c}),
it is understood that the following replacements are carried out,
\begin{eqnarray}
&~& [a(\bm{k}), a^{\dagger}(\bm{l})] = \delta^3(\bm{k} - \bm{l})
\to \{c(\bm{k}), c^{\dagger}(\bm{l})\} = \delta^3(\bm{k} - \bm{l})~,~~
\nonumber \\
&~& [b^{\dagger}(\bm{k}), b(\bm{l})] = -\delta^3(\bm{k} - \bm{l})
\to \{d^{\dagger}(\bm{k}), d(\bm{l})\} = - \delta^3(\bm{k} - \bm{l})~.
\label{replacements}
\end{eqnarray}
Note that the replacement $[b(\bm{l}), b^{\dagger}(\bm{k})] = \delta^3(\bm{k} - \bm{l})$
by $\{d(\bm{l}), d^{\dagger}(\bm{k})\} = \delta^3(\bm{k} - \bm{l})$
is incompatible with our anti-commutation relations (\ref{CCR-cd-c}).

The vacuum expectation values of the time ordered products are calculated as
\begin{eqnarray}
&~& \langle 0 |{\rm T}c_{\varphi}(x) c_{\varphi}^{\dagger}(y)| 0 \rangle 
= \langle 0 |(\theta(x_0- y_0)c_{\varphi}(x) c_{\varphi}^{\dagger}(y)
-\theta(y_0 - x_0)  c_{\varphi}^{\dagger}(y) c_{\varphi}(x)) | 0 \rangle 
\nonumber \\
&~& ~~~~~~~~~~~~~~~~~~~~~~~~~~~~~~~~~~~
= \int \frac{d^3k d^3l}{(2\pi)^3 \sqrt{2k_0~2l_0}} 
\left(\theta(x_0- y_0)\langle 0 |c(\bm{k}) c^{\dagger}(\bm{l})| 0 \rangle e^{-ikx+ily} \right.
\nonumber \\
&~& ~~~~~~~~~~~~~~~~~~~~~~~~~~~~~~~~~~~~~~~~~~~~~~~~~~~~~~~~~~~~~~~~~~~~~~~~~~~~ 
\left. - \theta(y_0 - x_0) \langle 0 | d(\bm{k}) d^{\dagger}(\bm{l})| 0 \rangle e^{ikx-ily}\right)
\nonumber \\
&~& ~~~~~~~~~~~~~~~~~~~~~~~~~~~~~~~~~~~ = \int \frac{d^3k}{(2\pi)^3 2k_0} 
\left(\theta(x_0- y_0)e^{-ik(x-y)} + \theta(y_0- x_0) e^{ik(x-y)}\right)
\nonumber \\
&~& ~~~~~~~~~~~~~~~~~~~~~~~~~~~~~~~~~~~ 
= \int \frac{d^4k}{(2\pi)^4} \frac{i e^{-ik(x-y)}}{k^2 - m^2 + i \varepsilon}
= i \varDelta_{\rm F}(x-y)~,~~ 
\label{Delta-F-c}\\
&~& \langle 0 |{\rm T}c_{\varphi}(x) c_{\varphi}(y)| 0 \rangle = 0~,~~
\langle 0 |{\rm T}c_{\varphi}^{\dagger}(x) c_{\varphi}^{\dagger}(y)| 0 \rangle = 0~,
\label{Tproduct-c}
\end{eqnarray}
where we use $\langle 0 |c(\bm{k}) c^{\dagger}(\bm{l})| 0 \rangle=\delta^3(\bm{k}-\bm{l})$,
$\langle 0 |d(\bm{k}) d^{\dagger}(\bm{l})| 0 \rangle=-\delta^3(\bm{k}-\bm{l})$
and so forth.
Hence, we obtain the same results as those in the case of the ordinary complex scalar field.

From $\{d(\bm{k}), d^{\dagger}(\bm{l})\} = -\delta^3(\bm{k}-\bm{l})$,
the negative norm states appear,
and the probability interpretation does not hold on.
Hence, it is difficult to construct a consistent quantum field theory
for a fermionic scalar field alone.

\subsection{Coexisting system of scalar fields}
\label{Coexisting system of scalar fields}

Now, let us consider the system that $(\varphi, \varphi^{\dagger})$ 
and $(c_{\varphi}, c_{\varphi}^{\dagger})$ coexist,
described by the Lagrangian density,
\begin{eqnarray}
\mathcal{L}_{\varphi, c_{\varphi}} = \partial_{\mu} \varphi^{\dagger} \partial^{\mu} \varphi 
- m^2 \varphi^{\dagger} \varphi +
\partial_{\mu} c_{\varphi}^{\dagger} \partial^{\mu} c_{\varphi} 
- m^2 c_{\varphi}^{\dagger} c_{\varphi}~.
\label{L-varphi-c}
\end{eqnarray}
From (\ref{H-ab-varphi}) and (\ref{H-cd-c}), the Hamiltonian is obtained as
\begin{eqnarray}
&~& H_{\varphi, c_{\varphi}} = \int d^3k k_0 
\left(a^{\dagger}(\bm{k}) a(\bm{k}) + b^{\dagger}(\bm{k}) b(\bm{k}) \right.
\left. + c^{\dagger}(\bm{k}) c(\bm{k}) - d^{\dagger}(\bm{k}) d(\bm{k})\right)~.
\label{H-abcd-varphi-c}
\end{eqnarray} 
Note that the sum of the zero-point energies vanishes due to the cancellation between 
contributions from  $(\varphi, \varphi^{\dagger})$ 
and $(c_{\varphi}, c_{\varphi}^{\dagger})$.

The eigenstates for $H_{\varphi, c_{\varphi}}$ are constructed by acting
the creation operators $a^{\dagger}(\bm{k})$, $b^{\dagger}(\bm{k})$,
$c^{\dagger}(\bm{k})$ and $d^{\dagger}(\bm{k})$ on the vacuum state $| 0 \rangle$.
This system also contains negative norm states, 
because the relation $\{d(\bm{k}), d^{\dagger}(\bm{l})\} = -\delta^3(\bm{k}-\bm{l})$
is imposed on.
In the same way as the coexisting system of harmonic oscillators,
it is shown that the system has fermionic symmetries 
and they rescue it from the difficulty.

The $\mathcal{L}_{\varphi, c_{\varphi}}$ is invariant under the fermionic transformations,
\begin{eqnarray}
\delta_{\rm F} \varphi = -\zeta c_{\varphi}~,~~\delta_{\rm F} \varphi^{\dagger} = 0~,~~ 
\delta_{\rm F} c_{\varphi}  = 0~,~~
\delta_{\rm F} c_{\varphi}^{\dagger} = \zeta \varphi^{\dagger}~
\label{delta-F-varphi}
\end{eqnarray}
and 
\begin{eqnarray}
\delta_{\rm F}^{\dagger} \varphi = 0~,~~
\delta_{\rm F}^{\dagger} \varphi^{\dagger} = \zeta^{\dagger} c_{\varphi}^{\dagger}~,~~
\delta_{\rm F}^{\dagger} c_{\varphi} = \zeta^{\dagger} \varphi~,~~
\delta_{\rm F}^{\dagger} c_{\varphi}^{\dagger} = 0~.
\label{delta-Fdagger-varphi}
\end{eqnarray}

From the above transformation properties, 
we see that ${\bm \delta}_{\rm F}$ and ${\bm \delta}_{\rm F}^{\dagger}$ are nilpotent,
i.e., ${Q_{\rm F}}^2 = 0$ and ${Q_{\rm F}^{\dagger}}^2 = 0$.
Here,
$Q_{\rm F}$ and $Q_{\rm F}^{\dagger}$ are the corresponding generators defined by
\begin{eqnarray}
\hspace{-1cm}&~& \zeta Q_{\rm F} 
\equiv \int d^3x \left[
\left(\frac{\partial \mathcal{L}_{\varphi, c_{\varphi}}}{\partial \dot{\varphi}}\right)_{\rm R} 
\delta_{\rm F} \varphi
+ \delta_{\rm F} c_{\varphi}^{\dagger} 
\left(\frac{\partial \mathcal{L}_{\varphi, c_{\varphi}}}{\partial \dot{c}_{\varphi}^{\dagger}}\right)_{\rm L}
\right]~,
\label{Q-F-def-varphi}\\
\hspace{-1cm}&~& Q_{\rm F}^{\dagger} \zeta^{\dagger} 
\equiv \int d^3x \left[
\delta_{\rm F}^{\dagger} \varphi^{\dagger}
 \left(\frac{\partial \mathcal{L}_{\varphi, c_{\varphi}}}{\partial \dot{\varphi}^{\dagger}}\right)_{\rm L} 
+
\left(\frac{\partial \mathcal{L}_{\varphi, c_{\varphi}}}{\partial \dot{c}_{\varphi}}\right)_{\rm R}
\delta_{\rm F}^{\dagger} c_{\varphi}
\right]~.
\label{Q-F-dagger-def-varphi}
\end{eqnarray}

We have the algebraic relation,
\begin{eqnarray}
\{Q_{\rm F}, Q_{\rm F}^{\dagger}\} = N_{\rm D}~,~
\label{QQdagger-varphi}
\end{eqnarray}
where $N_{\rm D}$ is the number operator defined by
\begin{eqnarray}
\hspace{-1cm}&~& N_{\rm D} \equiv - N_{\varphi} - N_{c_{\varphi}} 
 = \int d^3k \left(a^{\dagger}(\bm{k}) a(\bm{k}) - b^{\dagger}(\bm{k}) b(\bm{k}) \right.
\left. + c^{\dagger}(\bm{k}) c(\bm{k}) + d^{\dagger}(\bm{k}) d(\bm{k})\right) ~.
\label{ND-varphi}
\end{eqnarray}
$N_{\varphi}$ and $N_{c_{\varphi}}$ are generators
for $U(1)$ transformations of $\varphi$ and $c_{\varphi}$,
given by (\ref{Q-varphi}) and (\ref{Q-cvarphi}), respectively.
Note that the infinite constants in $N_{\varphi}$ and $N_{c_{\varphi}}$
are canceled out in $N_{\rm D}$ in the similar way as $H_{\varphi, c_{\varphi}}$.
The symmetry of our system is also equivalent to $OSp(2|2)$.

From (\ref{Q-F-def-varphi}) and (\ref{Q-F-dagger-def-varphi}),
the conserved fermionic charges $Q_{\rm F}$ and $Q_{\rm F}^{\dagger}$ are obtained by
\begin{eqnarray}
&~& Q_{\rm F} = \int d^3x \left( - \pi c_{\varphi} 
+  \varphi^{\dagger} \pi_{c_{\varphi}}^{\dagger}\right)
= - i \int d^3k \left(a^{\dagger}(\bm{k}) c(\bm{k}) - d^{\dagger}(\bm{k}) b(\bm{k})\right)~,~~
\label{QF-varphi}\\
&~& Q_{\rm F}^{\dagger} = 
\int d^3x \left(- c_{\varphi}^{\dagger} \pi^{\dagger}
+ \pi_{c_{\varphi}} \varphi\right)
=  i \int d^3k \left(c^{\dagger}(\bm{k}) a(\bm{k}) - b^{\dagger}(\bm{k}) d(\bm{k})\right)~.
\label{QF-dagger-varphi}
\end{eqnarray}
Then, the canonical momenta are transformed as,
\begin{eqnarray}
\delta_{\rm F} \pi = 0~,~~\delta_{\rm F} \pi^{\dagger} = -\zeta \pi_{c_{\varphi}}^{\dagger}~,~~ 
\delta_{\rm F} \pi_{c_{\varphi}}  = \zeta \pi~,~~
\delta_{\rm F} \pi_{c_{\varphi}}^{\dagger} = 0
\label{delta-F-pi}
\end{eqnarray}
and 
\begin{eqnarray}
\delta_{\rm F}^{\dagger} \pi = \zeta^{\dagger} \pi_{c_{\varphi}}~,~~
\delta_{\rm F}^{\dagger} \pi^{\dagger} = 0 ~,~~
\delta_{\rm F}^{\dagger} \pi_{c_{\varphi}} = 0~,~~
\delta_{\rm F}^{\dagger} \pi_{c_{\varphi}}^{\dagger} = -\zeta^{\dagger} \pi^{\dagger}~.
\label{delta-Fdagger-pi}
\end{eqnarray}

It is easily understood that $\mathcal{L}_{\varphi, c_{\varphi}}$ is invariant 
under the transformations (\ref{delta-F-varphi}) and (\ref{delta-Fdagger-varphi}),
from the nilpotency of ${\bm \delta}_{\rm F}$ and ${\bm \delta}_{\rm F}^{\dagger}$ 
and the relations,
\begin{eqnarray}
\mathcal{L}_{\varphi, c_{\varphi}} =  {\bm \delta}_{\rm F} \mathcal{R}_{\varphi, c_{\varphi}}
= {\bm \delta}_{\rm F}^{\dagger} \mathcal{R}_{\varphi, c_{\varphi}}^{\dagger}
= {\bm \delta}_{\rm F}  {\bm \delta}_{\rm F}^{\dagger} \mathcal{L}_{\varphi} 
= - {\bm \delta}_{\rm F}^{\dagger} {\bm \delta}_{\rm F} \mathcal{L}_{\varphi}~,
\label{delta-rel-varphi}
\end{eqnarray}
where $\mathcal{R}_{\varphi, c_{\varphi}}$ and $\mathcal{R}_{\varphi, c_{\varphi}}^{\dagger}$ 
are given by
\begin{eqnarray}
&~& \mathcal{R}_{\varphi, c_{\varphi}} =  \partial_{\mu} c_{\varphi}^{\dagger} \partial^{\mu} \varphi
- m^2 c_{\varphi}^{\dagger} \varphi ~,~~
\mathcal{R}_{\varphi, c_{\varphi}}^{\dagger} 
= \partial_{\mu} \varphi^{\dagger} \partial^{\mu} c_{\varphi}
 - m^2 \varphi^{\dagger} c_{\varphi}~.
\label{R-varphi}
\end{eqnarray}
The Hamiltonian density
$\mathcal{H}_{\varphi, c_{\varphi}}$ is written in the $Q_{\rm F}$ and $Q_{\rm F}^{\dagger}$
exact forms such that 
\begin{eqnarray}
\mathcal{H}_{\varphi, c_{\varphi}} 
=  i \left\{Q_{\rm F}, \tilde{\mathcal{R}}_{\varphi, c_{\varphi}}\right\}
= - i \left\{Q_{\rm F}^{\dagger}, \tilde{\mathcal{R}}_{\varphi, c_{\varphi}}^{\dagger}\right\}
=  \left\{Q_{\rm F},  \left\{Q_{\rm F}^{\dagger}, 
 \mathcal{H}_{\varphi}\right\}\right\}
= - \left\{Q_{\rm F}^{\dagger}, \left\{Q_{\rm F}, \mathcal{H}_{\varphi}\right\}\right\}~,
\label{H-varphi-c-exact}
\end{eqnarray}
where $\tilde{\mathcal{R}}_{\varphi, c_{\varphi}}$ 
and $\tilde{\mathcal{R}}_{\varphi, c_{\varphi}}^{\dagger}$ are given by
\begin{eqnarray}
\tilde{\mathcal{R}}_{\varphi, c_{\varphi}} =  \pi_{c_{\varphi}} \pi^{\dagger} 
+ \bm{\nabla} c_{\varphi}^{\dagger} \bm{\nabla} \varphi + m^2  c_{\varphi}^{\dagger} \varphi~,~~
\tilde{\mathcal{R}}_{\varphi, c_{\varphi}}^{\dagger} = \pi \pi_{c_{\varphi}}^{\dagger} 
+ \bm{\nabla} \varphi^{\dagger} \bm{\nabla} c_{\varphi} + m^2 \varphi^{\dagger} c_{\varphi}~.
\label{tildeR-varphi}
\end{eqnarray}

As in the case of the harmonic oscillators, 
negative norm states can be projected out by imposing 
the following subsidiary conditions on states,
\begin{eqnarray}
Q_{\rm F} |{\rm phys}\rangle = 0~,~~
Q_{\rm F}^{\dagger} |{\rm phys}\rangle = 0~,~~
N_{\rm D} |{\rm phys}\rangle = 0
\label{Phys-again}
\end{eqnarray}
or
\begin{eqnarray}
\tilde{Q}_1 |{\rm phys}\rangle = 0~,~~
\tilde{Q}_2 |{\rm phys}\rangle = 0~,~~
N_{\rm D} |{\rm phys}\rangle = 0~,
\label{Phys-Q12-again}
\end{eqnarray}
where $\tilde{Q}_1$ and $\tilde{Q}_2$ are defined by
\begin{eqnarray}
\tilde{Q}_1 \equiv Q_{\rm F} + Q_{\rm F}^{\dagger}~,~~
\tilde{Q}_2 \equiv i(Q_{\rm F} - Q_{\rm F}^{\dagger})~.
\label{Q12-again}
\end{eqnarray}
As a result, the theory becomes harmless but empty leaving the vacuum state alone.

Finally, we point out that the remedy of the system described by $\mathcal{L}_{c_{\varphi}}$ 
is not unique as in the case with the harmonic oscillator.
There is a possibility that the real and the imaginary part
of $c_{\varphi}$
are regarded as the Faddeev-Popov ghost field $c(x)$ 
and the anti-ghost field $\overline{c}(x)$, respectively.
Using $c_{\varphi}(x) = (c(x) + i \overline{c}(x))/\sqrt{2}$,
$\mathcal{L}_{c_{\varphi}}$ is rewritten by
\begin{eqnarray}
\mathcal{L}_{c_{\varphi}} 
= - i \partial_{\mu} \overline{c}(x) \partial^{\mu} {c}(x)
- i m^2  \overline{c}(x) {c}(x)~.
\label{L-c2}
\end{eqnarray}
As is well known,
in the presence of the gauge boson $A_{\mu}(x)$,
we can construct a consistent quantum theory 
containing massless scalar fields obeying anti-commutation relations.
Non-gauge model with a pair of hermitian scalar fields $(c(x), \overline{c}(x))$ 
and those BRST partners 
also has been constructed and studied~\cite{F2,F3}.

\section{Conclusions}
\label{Conclusions}

We have reexamined the connection between spin and statistics through
the quantization of a complex scalar field.
Starting from an ordinary Lagrangian density
and imposing the anti-commutation relations on the scalar field,
we have found that the difficulty stems from
not the ill-definiteness (or unboundedness) of the energy 
and the violation of the causality
but the appearance of states with negative norms.\footnote{
Recently, higher spin fields with abnormal commutation relations are
studied and the same features are obtained using explicit models in \cite{Toth}.
The models are different from ours in the following point.
The models in \cite{Toth} are constructed from a pair of complex fields
such as a pair of fermionic scalar fields or a pair of bosonic spinor fields
and their Lagrangian density is composed of the mixing terms of the pair.
In contrast, our model is constructed from a single fermionic scalar field 
or a single bosonic spinor field and the Lagrangian density
has the same form as that of an ordinary complex scalar field
or an ordinary Dirac spinor field.
If combined with our results,
it is not unreasonable to conjecture
that the positive norm condition is crucial to the spin-statistics theorem
in a wide class of models.
}$^,$\footnote{
In \cite{Ohta}, the connection of spin and statistics are examined
for massless fields in any number of space-time dimensions,
and it is concluded that
hermitian fields obeying abnormal relations like the Faddeev-Popov ghost fields
do not violate the microscopic causality, either.
}
These features also hold for the system with a spinor field imposing the commutation relations on.
As a by-product, we have constructed analytical mechanics
in the form with the manifestly hermitian property.

The fermionic scalar field (or a bosonic spinor field) cannot exist alone, 
because the probability of its discovery is negative and physically meaningless.
We have proposed that the system with a fermionic scalar field 
(or a bosonic spinor field)
becomes harmless by introducing an ordinary complex scalar field
(or an ordinary spinor field)
to form a doublet of fermionic symmetries,
although the system becomes empty leaving the vacuum state alone.
It is meaningful to construct an interacting model containing 
our coexisting system as a subsystem, after the example of 
the gauge fixing term and the Faddeev-Popov ghost term
in gauge theories.

Here, the following question arises from the physical point of view.
Even if there were a coexisting system with only unphysical modes,
is it physically meaningful or is it verified?
It is deeply connected to the question $\lq\lq$what is the physical reality?''
There is a possibility that unphysical particles leave behind a fingerprint 
relating symmetries based on the scenario that our world comes into existence
from unphysical world,
even if they did not give any dynamical effects on the physical sector
at the beginning~\cite{YK}.
It would be interesting to explore the physics concerning 
the reversal connection of spin and statistics
and the application to its phenomenology,
based on the above scenario.

\section*{Acknowledgments}
The author thanks Prof. T. Kugo for valuable discussions
and useful comments.
This work was supported in part by scientific grants from the Ministry of Education, Culture,
Sports, Science and Technology under Grant Nos.~22540272 and 21244036.

\appendix

\section{Differentiation, Hamiltonian
and analytical mechanics}

We present useful formulas of differentiation for variables.
For variables $A$ and $B$, the right-differentiation of $A B$ 
by a Grassmann variable $\theta_i$ is given by
\begin{eqnarray}
\left(\frac{\partial}{\partial \theta_i} (AB)\right)_{\rm R}
= A \left(\frac{\partial B}{\partial \theta_i}\right)_{\rm R}
+  (-)^{|B|} \left(\frac{\partial A}{\partial \theta_i}\right)_{\rm R} B~,
\label{AB-R}
\end{eqnarray}
where $|B|$ is the number representing the Grassmann parity of $B$, 
i.e., $|B|=1$ for the Grassmann odd $B$
and $|B|=0$ for the Grassmann even $B$.

The left-differentiation of $A B$ 
by $\theta_i$ is given by
\begin{eqnarray}
\left(\frac{\partial}{\partial \theta_i} (AB)\right)_{\rm L}
= \left(\frac{\partial A}{\partial \theta_i}\right)_{\rm L} B
+  (-)^{|A|} A \left(\frac{\partial B}{\partial \theta_i}\right)_{\rm L}~.
\label{AB-L}
\end{eqnarray}

We have the following relation between the right and the left-differentiation:
\begin{eqnarray}
\left[\left(\frac{\partial A}{\partial \theta_i}\right)_{\rm R}\right]^{\dagger}
= \left(\frac{\partial A^{\dagger}}{\partial \theta^{\dagger}_i}\right)_{\rm L}~.
\label{A-RL}
\end{eqnarray}
Actually, the hermitian conjugate of (\ref{AB-R}) is rewritten as
\begin{eqnarray}
\hspace{-1cm}&~& \left[\left(\frac{\partial}{\partial \theta_i} (AB)\right)_{\rm R}\right]^{\dagger}
= \left[\left(\frac{\partial B}{\partial \theta_i}\right)_{\rm R}\right]^{\dagger} A^{\dagger}
+ (-)^{|B|} B^{\dagger} \left[\left(\frac{\partial A}{\partial \theta_i}\right)_{\rm R}\right]^{\dagger}
= \left(\frac{\partial B^{\dagger}}{\partial \theta^{\dagger}_i}\right)_{\rm L} A^{\dagger}
+ (-)^{|B^{\dagger}|} B^{\dagger} 
\left(\frac{\partial A^{\dagger}}{\partial \theta^{\dagger}_i}\right)_{\rm L}
\nonumber \\
\hspace{-1cm}&~& ~~~~~~~~~~~~~~~~~~~~~~~~~~~~~~~
= \left(\frac{\partial}{\partial \theta^{\dagger}_i} (B^{\dagger} A^{\dagger})\right)_{\rm L}
= \left(\frac{\partial}{\partial \theta^{\dagger}_i} (AB)^{\dagger}\right)_{\rm L}~,
\label{AB-R-dagger}
\end{eqnarray}
where we use $(AB)^{\dagger} = B^{\dagger} A^{\dagger}$,
$|B| = |B^{\dagger}|$, (\ref{AB-L}) and (\ref{A-RL}).
This relation consists with (\ref{A-RL}). 

For any variable $z_n$ taking an ordinary or a Grassmann number, 
 (\ref{A-RL}) is generalized to
\begin{eqnarray}
\left[\left(\frac{\partial f(z_m, z^{\dagger}_m)}{\partial z_n}\right)_{\rm R}\right]^{\dagger}
= \left(\frac{\partial f^{\dagger}(z_m, z^{\dagger}_m)}{\partial z^{\dagger}_n}\right)_{\rm L}~.
\label{f-RL}
\end{eqnarray}

Let us develop analytical mechanics for the system 
with a set of variables $(Q_k, Q_k^{\dagger})$ containing bosonic and/or fermionic ones.
For the Lagrangian $L=L(Q_k, \dot{Q}_k, Q^{\dagger}_k, \dot{Q}_k^{\dagger})$,
we define the canonical momentum of $Q_k$ by
\begin{eqnarray}
P_k \equiv \left(\frac{\partial L}{\partial \dot{Q}_k}\right)_{\rm R}~.
\label{P}
\end{eqnarray}
Then the hermitian conjugate of $P_k$ is given by
\begin{eqnarray}
P_k^{\dagger} = \left[\left(\frac{\partial L}{\partial \dot{Q}_k}\right)_{\rm R}\right]^{\dagger}
= \left(\frac{\partial L^{\dagger}}{\partial \dot{Q}_k^{\dagger}}\right)_{\rm L}
= \left(\frac{\partial L}{\partial \dot{Q}_k^{\dagger}}\right)_{\rm L}~,
\label{P-dagger}
\end{eqnarray}
where we use (\ref{f-RL}) and $L^{\dagger} = L$.
Here, we adopt $P_k^{\dagger}$ defined in (\ref{P-dagger})
as the canonical momenta of $Q_k^{\dagger}$.
and then 
analytical mechanics can be constructed with the manifestly hermitian property
that the hermitian conjugate of canonical momentum for a variable is
just the canonical momentum for the hermitian conjugate of the variable.

Using $P_k$ and $P_k^{\dagger}$, the Hamiltonian is defined by
\begin{eqnarray}
H \equiv \sum_k  \left[\left(\frac{\partial L}{\partial \dot{Q}_k}\right)_{\rm R} \dot{Q}_k
+ \dot{Q}_k^{\dagger} \left(\frac{\partial L}{\partial \dot{Q}_k^{\dagger}}\right)_{\rm L}\right]
- L
= \sum_k \left(P_k \dot{Q}_k + \dot{Q}_k^{\dagger} P_k^{\dagger}\right) - L~,
\label{H-def}
\end{eqnarray}
where $H$ is hermitian by definition
and should be expressed using canonical variables
after $\dot{Q}_k$ and $\dot{Q}^{\dagger}_k$ are
obtained as functions of canonical ones.

Based on this definition, the variations of $L$ and $H$ are given by
\begin{eqnarray}
&~& \delta L = \sum_k  \left[\left(\frac{\partial L}{\partial {Q}_k}\right)_{\rm R} \delta Q_k
+ \left(\frac{\partial L}{\partial \dot{Q}_k}\right)_{\rm R} \delta \dot{Q}_k \right.
\left.
+ \delta {Q}_k^{\dagger} \left(\frac{\partial L}{\partial {Q}_k^{\dagger}}\right)_{\rm L}
+ \delta \dot{Q}_k^{\dagger} \left(\frac{\partial L}{\partial \dot{Q}_k^{\dagger}}\right)_{\rm L}
\right]~,~~
\label{delta-L}\\
&~& \delta H = \sum_k \left[\left(\frac{\partial H}{\partial {Q}_k}\right)_{\rm R} \delta {Q}_k
+ \delta {P}_k \left(\frac{\partial H}{\partial P_k}\right)_{\rm L} \right.
\left.
+ \delta {Q}_k^{\dagger} \left(\frac{\partial H}{\partial Q_k^{\dagger}}\right)_{\rm L}
+ \left(\frac{\partial H}{\partial {P}_k^{\dagger}}\right)_{\rm R} \delta {P}_k^{\dagger}
\right]~,
\label{delta-H}
\end{eqnarray}
where $L$ and $H$ are assumed not to contain the time variable $t$ explicitly.
From the variational principle, the following Hamilton's canonical equations
of motion are derived,
\begin{eqnarray}
\frac{d Q_k}{d t} = \left(\frac{\partial H}{\partial P_k}\right)_{\rm L}~,~~
\frac{d P_k}{d t} = - \left(\frac{\partial H}{\partial Q_k}\right)_{\rm R}~,~~
\frac{d Q_k^{\dagger}}{d t} = \left(\frac{\partial H}{\partial P_k^{\dagger}}\right)_{\rm R}~,~~
\frac{d P_k^{\dagger}}{d t} = - \left(\frac{\partial H}{\partial Q_k^{\dagger}}\right)_{\rm L}~.
\label{Hamilton-eq}
\end{eqnarray}
Then, the Hamilton equation for $F=F(Q_k, P_k, Q_k^{\dagger}, P_k^{\dagger})$ is written by
\begin{eqnarray}
\frac{d F}{d t} = \left\{F, H\right\}_{\rm PB}~,
\label{Hamilton-eq2}
\end{eqnarray}
where $\left\{f, g\right\}_{\rm PB}$ is the Poisson bracket defined by
\begin{eqnarray}
\hspace{-1cm}&~& \left\{f, g\right\}_{\rm PB} \equiv 
\sum_{k} \left[\left(\frac{\partial f}{\partial Q_k}\right)_{\rm R}
 \left(\frac{\partial g}{\partial P_k}\right)_{\rm L}
- (-)^{|Q_k|} \left(\frac{\partial f}{\partial P_k}\right)_{\rm R}
 \left(\frac{\partial g}{\partial Q_k}\right)_{\rm L} \right.
\nonumber \\
\hspace{-1cm}&~& ~~~~~~~~~~~~~~~
\left. +  (-)^{|Q_k|} \left(\frac{\partial f}{\partial Q_k^{\dagger}}\right)_{\rm R}
\left(\frac{\partial g}{\partial P_k^{\dagger}}\right)_{\rm L}
- \left(\frac{\partial f}{\partial P_k^{\dagger}}\right)_{\rm R}
\left(\frac{\partial g}{\partial Q_k^{\dagger}}\right)_{\rm L}
\right]~.
\label{Poisson}
\end{eqnarray}
We see that (\ref{Hamilton-eq}) is derived from (\ref{Hamilton-eq2})
using the relations,
\begin{eqnarray}
\left(\frac{\partial H}{\partial {Q}_k}\right)_{\rm L}
= (-)^{|Q_k|} \left(\frac{\partial H}{\partial {Q}_k}\right)_{\rm R}~,~~
\left(\frac{\partial H}{\partial {P}^{\dagger}_k}\right)_{\rm R}
= (-)^{|Q_k|} \left(\frac{\partial H}{\partial {P}^{\dagger}_k}\right)_{\rm L}~.
\label{H-RL}
\end{eqnarray}
Note that $(-)^{|Q_k|} = (-)^{|P_k|} = (-)^{|Q^{\dagger}_k|} = (-)^{|P^{\dagger}_k|}$
and $(-)^{2|Q_k|}=1$.

We see that the following relations concerning the above Poisson bracket hold on:
\begin{eqnarray}
\hspace{-1cm}&~& \left\{f, g\right\}_{\rm PB} = (-)^{|f||g|+1} \left\{g, f\right\}_{\rm PB}~,
\label{PB-1}\\
\hspace{-1cm}&~& \left\{f, \alpha g + \beta h\right\}_{\rm PB} 
= \alpha \left\{f, g\right\}_{\rm PB} + \beta  \left\{f, h\right\}_{\rm PB}~,
\label{PB-2}\\
\hspace{-1cm}&~& \left\{f g, h\right\}_{\rm PB} = f \left\{g, h\right\}_{\rm PB} 
+ (-)^{|g||h|} \left\{f, h\right\}_{\rm PB} g~,
\label{PB-3}\\
\hspace{-1cm}&~& (-)^{|h||f|} \left\{ \left\{f, g\right\}_{\rm PB}, h\right\}_{\rm PB}
+ (-)^{|f||g|} \left\{ \left\{g, h\right\}_{\rm PB}, f\right\}_{\rm PB}
+ (-)^{|g||h|} \left\{ \left\{h, f\right\}_{\rm PB}, g\right\}_{\rm PB} = 0~,
\label{PB-4}
\end{eqnarray}
where $\alpha$ and $\beta$ are quantities irrelevant to canonical variables,
and the last relation is the Jacobi identity.

The canonical quantization is carried out by regarding variables
as operators and replacing the Poisson bracket into
the commutation relation for bosonic variables or the anti-commutation relation
for fermionic variables
such that
\begin{eqnarray}
 \left\{f, g\right\}_{\rm PB}
\to \frac{1}{i \hbar} [f, g]~~~~ {\rm or} ~~~~ \frac{1}{i \hbar} \{f, g\} ~.
\label{Quantization}
\end{eqnarray}

Let $L$ be invariant under the transformation,
\begin{eqnarray}
\delta Q_k = i [\epsilon N, Q_k]~,~~
\delta Q_k^{\dagger} = i [\epsilon N, Q_k^{\dagger}]~,
\label{U(1)-Q}
\end{eqnarray}
where $\epsilon$ is an infinitesimal real number
and $N$ is the conserved Noether charge defined by
\begin{eqnarray}
&~& \epsilon N 
\equiv \sum_k 
\left[\left(\frac{\partial L}{\partial \dot{Q}_k}\right)_{\rm R} \delta Q_k
+ \delta Q_k^{\dagger} \left(\frac{\partial L}{\partial \dot{Q}_k^{\dagger}}\right)_{\rm L}\right]
= \sum_k 
\left(P_k \delta Q_k + \delta Q_k^{\dagger} P_k^{\dagger}\right)~,
\label{N-def}
\end{eqnarray}
where $N$ is also hermitian by definition.

For the Lagrangian density 
$\mathcal{L} 
=\mathcal{L}(\phi^a, \partial_{\mu} \phi^a, \phi^{a \dagger}, \partial_{\mu} \phi^{a \dagger})$
with $(\phi^a, \phi^{a \dagger})$
containing bosonic and/or fermionic variables,
let $\mathcal{L}$ be invariant under the transformation (irrelevant to the space-time)
$\phi^a \to \phi^a + \delta \phi^a$ 
and $\phi^{a \dagger} \to \phi^{a \dagger} + \delta \phi^{a \dagger}$.
Then, the Noether current $j^{\mu}$ is defined by
\begin{eqnarray}
\epsilon j^{\mu}
\equiv \sum_k 
\left[\left(\frac{\partial \mathcal{L}}{\partial \partial_{\mu} \phi^a}\right)_{\rm R} \delta \phi^a
+ \delta \phi^{a \dagger} 
\left(\frac{\partial \mathcal{L}}{\partial_{\mu} \phi^{a \dagger}}\right)_{\rm L}\right]
\label{j-def}
\end{eqnarray}
and is subject to the conservation law such as $\partial_{\mu} j^{\mu} = 0$.

\section{Bosonic spinor field}

We study the system of a bosonic spinor field
(a spinor field
imposing the commutation relations on variables), and clarify the difficulty on quantization.

Let us take the system described by the Lagrangian density,
\begin{eqnarray}
\mathcal{L}_{c_{\psi}} = i \overline{c}_{\psi} \gamma^{\mu} \partial_{\mu} c_{\psi}
-m  \overline{c}_{\psi}c_{\psi}~,
\label{L-cpsi}
\end{eqnarray}
where $c_{\psi}$ is a spinor field taking complex numbers,
$\overline{c}_{\psi} \equiv c_{\psi}^{\dagger} \gamma^0$
and $\gamma^{\mu}$ are the gamma matrices satisfying 
$\{\gamma^{\mu}, \gamma^{\nu}\} = 2 \eta^{\mu\nu}$.
The Euler-Lagrange equations for $c_{\psi}$ and $\overline{c}_{\psi}$
are given by
\begin{eqnarray}
 \overline{c}_{\psi} \left(i \gamma^{\mu} \overleftarrow{\partial}_{\mu} + m\right) = 0~,~~
\left(i \gamma^{\mu} \partial_{\mu} - m\right) {c}_{\psi} = 0~,
\label{D-eq-cpsi}
\end{eqnarray}
respectively.
Here and hereafter, we use $\mathcal{L}_{c_{\psi}}$
in place of the hermitian one,
\begin{eqnarray}
\mathcal{L}_{c_{\psi}}^0 = \frac{i}{2} \left(\overline{c}_{\psi} \gamma^{\mu} {\partial}_{\mu} c_{\psi}
- \partial_{\mu} \overline{c}_{\psi} \gamma^{\mu} c_{\psi}\right)
-m  \overline{c}_{\psi}c_{\psi}~,
\label{L-cpsi-hermite}
\end{eqnarray}
because the same conclusions are obtained easier.

The canonical conjugate momentum of $c_{\psi}$ is given by
\begin{eqnarray}
\pi_{c_{\psi}} \equiv \left(\frac{\partial \mathcal{L}_{c_{\psi}}}{\partial \dot{c}_{\psi}}\right)_{\rm R}
=  i \overline{c}_{\psi} \gamma^0 = i c_{\psi}^{\dagger} ~.
\label{pi-cpsi}
\end{eqnarray}

By solving (\ref{D-eq-cpsi}) and (\ref{pi-cpsi}), we obtain the solutions,
\begin{eqnarray}
&~& c_{\psi}(x) = \int \frac{d^3k}{\sqrt{(2\pi)^3 2k_0}}\sum_{s}
\left(\tilde{c}(\bm{k}, s) u(\bm{k}, s) e^{-i k x} \right.
 \left. + \tilde{d}^{\dagger}(\bm{k}, s) v(\bm{k}, s)e^{i k x}\right)~,
\label{cpsi-sol}\\
&~& \pi_{c_{\psi}}(x) = i \int \frac{d^3k}{\sqrt{(2\pi)^3 2k_0}}\sum_{s}
\left(\tilde{c}^{\dagger}(\bm{k}, s) u^{\dagger}(\bm{k}, s) e^{i k x}  \right.
\left. 
+ \tilde{d} (\bm{k}, s) v^{\dagger}(\bm{k}, s) e^{-i k x}\right)~,
\label{pi-cpsi-sol}
\end{eqnarray}
where $s$ represents the spin state,
and $u(\bm{k}, s)$ and $v(\bm{k}, s)$ are Dirac spinors on the momentum space.
They satisfy the relations,
\begin{eqnarray}
\sum_s u(\bm{k}, s) \overline{u}(\bm{k}, s) = \Slashk + m~,~~
\sum_s v(\bm{k}, s) \overline{v}(\bm{k}, s) = \Slashk - m~,
\label{uv}
\end{eqnarray}
where $\overline{u}(\bm{k}, s) \equiv {u}^{\dagger}(\bm{k}, s)\gamma^0$,
$\overline{v}(\bm{k}, s) \equiv {v}^{\dagger}(\bm{k}, s)\gamma^0$
and $\Slashk = \gamma^{\mu} k_{\mu}$.

Using (\ref{pi-cpsi}), the Hamiltonian density is obtained as
\begin{eqnarray}
\mathcal{H}_{c_{\psi}} = \pi_{c_{\psi}} \dot{c}_{\psi} - \mathcal{L}_{c_{\psi}}
= - i \sum_{i=1}^{3} \overline{c}_{\psi} \gamma^i \partial_i c_{\psi} + m  \overline{c}_{\psi}c_{\psi}~.
\label{H-cpsi}
\end{eqnarray}

Let us quantize the system regarding variables as operators
and imposing the following commutation relations 
on $(c_{\psi}, \pi_{c_{\psi}})$,
\begin{eqnarray}
\hspace{-1cm}&~& [c_{\psi}^{\alpha}(\bm{x}, t), \pi_{c_{\psi}}^{\beta}(\bm{y}, t)]
 = i \delta^{\alpha \beta}\delta^3(\bm{x}-\bm{y})~,~~ 
\nonumber \\
\hspace{-1cm}&~& 
[c_{\psi}^{\alpha}(\bm{x}, t), c_{\psi}^{\beta}(\bm{y}, t)] = 0~,~~
[\pi_{c_{\psi}}^{\alpha}(\bm{x}, t), \pi_{c_{\psi}}^{\beta}(\bm{y}, t)] = 0~,
\label{CCR-cpsi}
\end{eqnarray}
where $\alpha$ and $\beta$ are spinor indices.
Or equivalently, for operators $\tilde{c}(\bm{k}, s)$, $\tilde{d}^{\dagger}(\bm{k}, s)$, 
$\tilde{c}^{\dagger}(\bm{k}, s)$ and $\tilde{d}(\bm{k}, s)$,
the following commutation relations are imposed on,
\begin{eqnarray}
&~& [\tilde{c}(\bm{k}, s), \tilde{c}^{\dagger}(\bm{l}, s')] = \delta_{s s'} \delta^3(\bm{k}-\bm{l})~,~~ 
[\tilde{d}(\bm{k}, s), \tilde{d}^{\dagger}(\bm{l}, s')] = - \delta_{s s'} \delta^3(\bm{k}-\bm{l})~,~~ 
\nonumber \\
&~& [\tilde{c}(\bm{k}, s), \tilde{d}(\bm{l}, s')] = 0~,~~ 
[\tilde{c}^{\dagger}(\bm{k}, s), \tilde{d}^{\dagger}(\bm{l}, s')] = 0~,~~
[\tilde{c}(\bm{k}, s), \tilde{d}^{\dagger}(\bm{l}, s')] = 0~,~~ 
\nonumber \\
&~&[\tilde{c}^{\dagger}(\bm{k}, s), \tilde{d}(\bm{l}, s')] = 0~,~~
 [\tilde{c}(\bm{k}, s), \tilde{c}(\bm{l}, s')] = 0~,~~ 
[\tilde{c}^{\dagger}(\bm{k}, s), \tilde{c}^{\dagger}(\bm{l}, s')] = 0~,~~
\nonumber \\
&~& [\tilde{d}(\bm{k}, s), \tilde{d}(\bm{l}, s')] = 0~,~~ 
[\tilde{d}^{\dagger}(\bm{k}, s), \tilde{d}^{\dagger}(\bm{l}, s')] = 0~.
\label{CCR-cd-cpsi}
\end{eqnarray}

Using (\ref{R-C}), (\ref{H-cpsi}), (\ref{CCR-cpsi}) and the Heisenberg equation, 
(\ref{D-eq-cpsi}) and (\ref{pi-cpsi}) are derived
where the Hamiltonian is given by $H_{c_{\psi}} = \int \mathcal{H}_{c_{\psi}} d^3x$.

By inserting (\ref{cpsi-sol}) and (\ref{pi-cpsi-sol}) into (\ref{H-cpsi}), 
the Hamiltonian $H_{c_{\psi}}$ is written by
\begin{eqnarray}
\hspace{-1cm}&~& H_{c_{\psi}} = \int d^3k \sum_{s} k_0 
\left(\tilde{c}^{\dagger}(\bm{k}, s) \tilde{c}(\bm{k}, s) 
- \tilde{d}(\bm{k}, s) \tilde{d}^{\dagger}(\bm{k}, s)\right)
\nonumber \\
\hspace{-1cm}&~& ~~~~~~~~ = \int d^3k \sum_{s} k_0 
\left(\tilde{c}^{\dagger}(\bm{k}, s) \tilde{c}(\bm{k}, s) 
- \tilde{d}^{\dagger}(\bm{k}, s) \tilde{d}(\bm{k}, s)\right)
+ \int \frac{d^3k d^3x}{(2\pi)^3} \sum_{s} k_0~.
\label{H-cd-cpsi}
\end{eqnarray}
Let us define the ground state $| 0 \rangle$ as the state that 
satisfies $\tilde{c}(\bm{k}, s) |0 \rangle = 0$ and $\tilde{d}(\bm{k}, s) |0 \rangle = 0$.
Then, the eigenstates and eigenvalues of $H_{c_{\psi}}$ are given by
\begin{eqnarray}
&~& \int d^3k_1 d^3k_2 \cdots d^3k_{n_a} d^3l_1 d^3l_2 \cdots d^3l_{n_b}
f_1(\bm{k}_1) f_2(\bm{k}_2) \cdots f_{n_a}(\bm{k}_{n_a}) 
g_1(\bm{l}_1) g_2(\bm{l}_2) \cdots g_{n_b}(\bm{l}_{n_b}) 
\nonumber \\
&~& ~~~ \cdot \tilde{c}^{\dagger}(\bm{k}_1, s_1) \tilde{c}^{\dagger}(\bm{k}_2, s_2) \cdots 
\tilde{c}^{\dagger}(\bm{k}_{n_a}, s_{n_a})
\tilde{d}^\dagger(\bm{l}_1, s'_1) \tilde{d}^{\dagger}(\bm{l}_2, s'_2) \cdots 
\tilde{d}^{\dagger}(\bm{l}_{n_b}, s'_{n_b}) |0\rangle~,~~
\label{H-cpsi-states}\\
&~& E= k_{10} + k_{20} + \cdots + k_{n_a 0} + l_{10} + l_{20} + \cdots + l_{n_b 0}~,
\label{E-cpsi}
\end{eqnarray}
where we subtract an infinite constant corresponding to the sum of the zero-point energies.
From (\ref{E-cpsi}), we find that the positivity of energy holds on,
although the commutation relations are imposed on
the spinor field
and the negative sign appears in front of $\tilde{d}^{\dagger}(\bm{k}, s) \tilde{d}(\bm{k}, s)$
in $H_{c_{\psi}}$.
Note that the negative sign also exists in
$[\tilde{d}(\bm{k}, s), \tilde{d}^{\dagger}(\bm{l}, s')] = - \delta_{s s'} \delta^3(\bm{k}-\bm{l})$
and it guarantees the positivity of energy.

$\mathcal{L}_{c_{\psi}}$ is invariant under the $U(1)$ transformation,
\begin{eqnarray}
\hspace{-0.6cm}\delta c_{\psi} = i [\epsilon N_{c_{\psi}}, c_{\psi}] = i \epsilon c_{\psi}~,~~
\delta c_{\psi}^{\dagger} = i [\epsilon N_{c_{\psi}}, c_{\psi}^{\dagger}]
= - i \epsilon c_{\psi}^{\dagger}~,
\label{U(1)-cpsi}
\end{eqnarray}
where $N_{c_{\psi}}$ is the conserved $U(1)$ charge given by
\begin{eqnarray}
\hspace{-1cm}&~& N_{c_{\psi}} 
= - \int d^3x c^{\dagger}_{\psi} c_{\psi}  
= -\int d^3k \sum_s
\left(\tilde{c}^{\dagger}(\bm{k}, s) \tilde{c}(\bm{k}, s) 
+\tilde{d}(\bm{k}, s)  \tilde{d}^{\dagger}(\bm{k}, s) \right)
\nonumber \\
\hspace{-1cm}&~& ~~~~~~~~ = - \int d^3k \sum_s
\left(\tilde{c}^{\dagger}(\bm{k}, s) \tilde{c}(\bm{k}, s) 
+ \tilde{d}^{\dagger}(\bm{k}, s) \tilde{d}(\bm{k}, s)\right)~,
\label{Q-cpsi}
\end{eqnarray}
where we subtract an infinite constant.
We find that the $U(1)$ charge of particle corresponding $\tilde{d}^{\dagger}(\bm{k}, s) | 0 \rangle$
is opposite to that corresponding $\tilde{c}^{\dagger}(\bm{k}, s) | 0 \rangle$
from $[\tilde{d}(\bm{k}, s), \tilde{d}^{\dagger}(\bm{l}, s')]
= - \delta_{s s'} \delta^3(\bm{k}-\bm{l})$
and $[\tilde{c}(\bm{k}, s), \tilde{c}^{\dagger}(\bm{l}, s')]
= \delta_{s s'} \delta^3(\bm{k}-\bm{l})$.
Hence, $\tilde{c}(\bm{k}, c)$ and $\tilde{d}^{\dagger}(\bm{k}, s)$ 
in $c_{\psi}(x)$ are regarded as
the annihilation operator of particle and the creation operator of antiparticle, respectively.
$\tilde{c}^{\dagger}(\bm{k}, c)$ and $\tilde{d}(\bm{k}, s)$ 
in $\pi_{c_\psi}(x)$ are regarded as
the creation operator of particle and the annihilation operator of antiparticle, respectively.

The 4-dimensional commutation relations are calculated as
\begin{eqnarray}
&~& [c_{\psi}^{\alpha}(x), \overline{c}_{\psi}^{\beta}(y)] 
= \int \frac{d^3k d^3l}{(2\pi)^3 \sqrt{2k_0~2l_0}} \sum_{s, s'}
\left([\tilde{c}(\bm{k}, s), \tilde{c}^{\dagger}(\bm{l}, s')] 
u^{\alpha}(\bm{k}, s) \overline{u}^{\beta}(\bm{l}, s') e^{-ikx+ily}
\right. 
\nonumber \\
&~& ~~~~~~~~~~~~~~~~~~~~~~~~~~~~~~~~~~~~~~~~~~~~~~~~~~~~~~~~~~~~~~~~~~~~~~~ \left. 
+ [\tilde{d}^{\dagger}(\bm{k}, s), \tilde{d}(\bm{l}, s')] 
v^{\alpha}(\bm{k}, s) \overline{v}^{\beta}(\bm{l}, s') e^{ikx-ily}\right)
\nonumber \\
&~& ~~~~~~~~~~~~~~~~~~~~~~~~~~~
= \int \frac{d^3k}{(2\pi)^3 2k_0} \sum_{s}
\left(u^{\alpha}(\bm{k}, s) \overline{u}^{\beta}(\bm{k}, s) e^{-ik(x-y)}
+ v^{\alpha}(\bm{k}, s) \overline{v}^{\beta}(\bm{k}, s) e^{ik(x-y)}\right)
\nonumber \\
&~& ~~~~~~~~~~~~~~~~~~~~~~~~~~~ 
= \left(i \gamma^{\mu} \partial_{\mu} + m\right)^{\alpha\beta}
 \int \frac{d^3k}{(2\pi)^3 2k_0} \left(e^{-ik(x-y)} - e^{ik(x-y)}\right)
\nonumber \\
&~& ~~~~~~~~~~~~~~~~~~~~~~~~~~~ 
=\left(i \gamma^{\mu} \partial_{\mu} + m\right)^{\alpha\beta} i \varDelta(x-y)
\equiv i S^{\alpha\beta}(x-y)~,~~ 
\label{S}\\
&~& [c_{\psi}^{\alpha}(x), c_{\psi}^{\beta}(y)] = 0~,~~
[\overline{c}_{\psi}^{\alpha}(x), \overline{c}_{\psi}^{\beta}(y)] = 0~,
\label{4D-CCR-cpsi}
\end{eqnarray}
where we use $[\tilde{c}(\bm{k}, s), \tilde{c}^{\dagger}(\bm{l}, s')] 
= \delta_{s s'} \delta^3(\bm{k}-\bm{l})$ and
$[\tilde{d}(\bm{k}, s), \tilde{d}^{\dagger}(\bm{l}, s')] = - \delta_{s s'} \delta^3(\bm{k}-\bm{l})$.
We find that two fields separated by a space-like interval commute with each other
from the relation $\varDelta(x-y) = 0$ for $(x-y)^2 < 0$,
and hence the microscopic causality also holds on.

The vacuum expectation values of the time ordered products are calculated as
\begin{eqnarray}
\hspace{-0.8cm}&~& \langle 0 |{\rm T}c_{\psi}^{\alpha}(x) \overline{c}_{\psi}^{\beta}(y)| 0 \rangle 
= \langle 0 |(\theta(x_0 - y_0) c_{\psi}^{\alpha}(x) \overline{c}_{\psi}^{\beta}(y)
 + \theta(y_0 - x_0) \overline{c}_{\psi}^{\beta}(y) c_{\psi}^{\alpha}(x))| 0 \rangle 
\nonumber \\
\hspace{-0.8cm}&~& ~~~~~~~~~~~~~~~~~~~~~~~~~~~~~~~~~~~ = \int \frac{d^3k d^3l}{(2\pi)^3 \sqrt{2k_0~2l_0}} 
\nonumber \\
\hspace{-0.8cm}&~& ~~~~~~~~~~~~~~~~~~~~~~~~~~~~~~~~~~~~~~~~~~~
\sum_{s, s'}
\left(\theta(x_0- y_0)\langle 0 |\tilde{c}(\bm{k}, s) \tilde{c}^{\dagger}(\bm{l}, s')| 0 \rangle
u^{\alpha}(\bm{k}, s) \overline{u}^{\beta}(\bm{l}, s') e^{-ikx+ily} \right.
\nonumber \\
\hspace{-0.8cm}&~& ~~~~~~~~~~~~~~~~~~~~~~~~~~~~~~~~~~~~~~~~~~~~~~~~~~~~~~~~~~~~~~
\left. + \theta(y_0 - x_0) \langle 0 | \tilde{d}(\bm{k}, s) \tilde{d}^{\dagger}(\bm{l}, s')| 0 \rangle 
v^{\alpha}(\bm{k}, s) \overline{v}^{\beta}(\bm{l}, s') e^{ikx-ily}\right)
\nonumber \\
\hspace{-0.8cm}&~& ~~~~~~~~~~~~~~~~~~~~~~~~~~~~~~~~~~~ = \int \frac{d^3k}{(2\pi)^3 2k_0} 
\sum_{s}
\left(\theta(x_0- y_0) u^{\alpha}(\bm{k}, s) \overline{u}^{\beta}(\bm{k}, s) e^{-ik(x-y)} \right.
\nonumber \\
\hspace{-0.8cm}&~& ~~~~~~~~~~~~~~~~~~~~~~~~~~~~~~~~~~~~~~~~~~~~~~~~~~~~~~~~~~~~~~~~~~~~~~~
\left. - \theta(y_0 - x_0) v^{\alpha}(\bm{k}, s) \overline{v}^{\beta}(\bm{k}, s) e^{ik(x-y)}\right)
\nonumber \\
\hspace{-0.8cm}&~& ~~~~~~~~~~~~~~~~~~~~~~~~~~~~~~~~~~~ = 
\left(i \gamma^{\mu} \partial_{\mu} + m\right)^{\alpha\beta} \int \frac{d^3k}{(2\pi)^3 2k_0} 
\left(\theta(x_0- y_0)e^{-ik(x-y)} + \theta(y_0- x_0) e^{ik(x-y)}\right)
\nonumber \\
\hspace{-0.8cm}&~& ~~~~~~~~~~~~~~~~~~~~~~~~~~~~~~~~~~~ = 
\left(i \gamma^{\mu} \partial_{\mu} + m\right)^{\alpha\beta} i \varDelta_{\rm F}(x-y)
\nonumber \\
\hspace{-0.8cm}&~& ~~~~~~~~~~~~~~~~~~~~~~~~~~~~~~~~~~~ 
= \int \frac{d^4k}{(2\pi)^4} \left(\frac{i e^{-ik(x-y)}}{\Slashk - m + i \varepsilon}\right)^{\alpha\beta}
= i S_{\rm F}^{\alpha\beta}(x-y)~,~~ 
\label{S-F}\\
\hspace{-0.8cm}&~& \langle 0 |{\rm T}c_{\psi}(x) c_{\psi}(y)| 0 \rangle = 0~,~~
\langle 0 |{\rm T}\overline{c}_{\psi}(x) \overline{c}_{\psi}(y)| 0 \rangle = 0~,
\label{Tproduct-cpsi}
\end{eqnarray}
where we use $\langle 0 |\tilde{c}(\bm{k}, s) \tilde{c}^{\dagger}(\bm{l}, s')| 0 \rangle
= \delta_{s s'} \delta^3(\bm{k}-\bm{l})$
and $\langle 0 | \tilde{d}(\bm{k}, s) \tilde{d}^{\dagger}(\bm{l}, s')| 0 \rangle
= - \delta_{s s'} \delta^3(\bm{k}-\bm{l})$, and
$S_{\rm F}^{\alpha\beta}(x-y)$ is the Feynman propagator for spinors.
Hence we obtain the same results as those in the case of the ordinary
Dirac spinor field.

From $[\tilde{d}(\bm{k}, s), \tilde{d}^{\dagger}(\bm{l}, s')]
= - \delta_{s s'} \delta^3(\bm{k}-\bm{l})$,
we find that the negative norm states appear
and the probability interpretation does not hold on.\footnote{
If we take $\tilde{d}^{\dagger}(\bm{k}, s)|0 \rangle = 0$ 
in place of $\tilde{d}(\bm{k}, s)|0 \rangle = 0$,
negative norm states do not appear but the positivity of energy is ruined.
This possibility might not be reasonable
because $\tilde{d}(\bm{k}, s)$ ($\tilde{d}^{\dagger}(\bm{k}, s)$) is normally interpreted
as the annihilation (creation) operator of antiparticle and hence it is natural to choose
the condition $\tilde{d}(\bm{k}, s)|0 \rangle = 0$.
}
Hence, it is difficult to construct a consistent quantum field theory
for a spinor field obeying the commutation relations alone.
This difficulty is also recovered by introducing an ordinary spinor field $\psi$
obeying the anti-commutation relations.
Concretely, for the system described by the Lagrangian density~\cite{YK},
\begin{eqnarray}
\mathcal{L}_{\psi, c_{\psi}} 
= i \overline{\psi} \gamma^{\mu} \partial_{\mu} \psi
-m \overline{\psi}\psi
+ i \overline{c}_{\psi} \gamma^{\mu} \partial_{\mu} c_{\psi}
-m  \overline{c}_{\psi}c_{\psi}~,
\label{L-psi-cpsi}
\end{eqnarray}
the theory becomes harmless but empty leaving the vacuum state alone,
assisted by fermionic symmetries, that is, the invariance under the transformations,
\begin{eqnarray}
\delta_{\rm F} \psi = \zeta c_{\psi}~,~~\delta_{\rm F} \psi^{\dagger} = 0~,~~ 
\delta_{\rm F} c_{\psi}  = 0~,~~
\delta_{\rm F} c_{\psi}^{\dagger} = \zeta \psi^{\dagger}~,~~
\label{delta-F-psi}
\end{eqnarray}
and 
\begin{eqnarray}
\delta_{\rm F}^{\dagger} \psi = 0~,~~
\delta_{\rm F}^{\dagger} \psi^{\dagger} = \zeta^{\dagger} c_{\psi}^{\dagger}~,~~
\delta_{\rm F}^{\dagger} c_{\psi} = \zeta^{\dagger} \psi~,~~
\delta_{\rm F}^{\dagger} c_{\psi}^{\dagger} = 0~.
\label{delta-Fdagger-psi}
\end{eqnarray}

\end{document}